\documentclass[preprint2]{proto}
\newcommand{\refs}{\par\noindent\hangindent=1pc\hangafter=1}
\begin{document}

\title{\textbf{\LARGE X-ray Properties of Young Stars and Stellar Clusters}}

\author {\textbf{\large Eric Feigelson and Leisa Townsley}}
\affil{\small\em Pennsylvania State University}

\author{\textbf{\large Manuel G\"udel}}
\affil{\small\em Paul Scherrer Institute}

\author{\textbf{\large Keivan Stassun}}
\affil{\small\em Vanderbilt University} 

\begin{abstract}
\baselineskip = 11pt
\leftskip = 0.65in
\rightskip = 0.65in
\parindent=1pc
{\small 

Although the environments of star and planet formation are 
thermodynamically cold, substantial X-ray emission from $10-100$ MK 
plasmas is present.  In low mass pre-main sequence stars, X-rays are 
produced by violent magnetic reconnection flares.  In high mass O 
stars, they are produced by wind shocks on both stellar and parsec
scales. The recent {\it Chandra} Orion Ultradeep Project, {\it 
XMM-Newton} Extended Survey of Taurus, and {\it Chandra} studies of 
more distant high-mass star forming regions reveal a wealth of X-ray 
phenomenology and astrophysics.  X-ray flares mostly resemble 
solar-like magnetic activity from multipolar surface fields, although 
extreme flares may arise in field lines extending to the 
protoplanetary disk.  Accretion plays a secondary role.  Fluorescent 
iron line emission and absorption in inclined disks demonstrate that 
X-rays can efficiently illuminate disk material. The consequent 
ionization of disk gas and irradiation of disk solids addresses a 
variety of important astrophysical issues of disk dynamics, planet 
formation, and meteoritics.  New observations of massive star forming 
environments such as M 17, the Carina Nebula and 30 Doradus show 
remarkably complex X-ray morphologies including the low-mass stellar 
population, diffuse X-ray flows from blister HII regions, and 
inhomogeneous superbubbles.  X-ray astronomy is thus providing 
qualitatively new insights into star and planet formation.  
\\~\\~\\~}

\end{abstract}

{\textbf{1. INTRODUCTION}}

Star and planet formation is generally viewed as a hydrodynamic 
process involving gravitational collapse of interstellar material at 
low temperatures, 10--100~K in molecular cloud cores and 100--1500~K 
in protoplanetary disks.  If thermodynamical equilibrium holds, this 
material should be neutral except in localized H{\sc II} regions 
where the bolometric ultraviolet emission from massive O star 
photoionization is present.  However, stars have turned out to be 
sources of intense X-rays at almost every stage of early formation 
and evolution, from low-mass brown dwarfs to massive O stars, to an 
extent that the stellar environment is ionized and heated (beyond 
effects due to ultraviolet radiation) out to considerable distances 
and thus made accessible to magnetic fields. 

X-ray observations reveal the presence of highly-ionized plasma with 
temperatures of $10^7 - 10^8$~K.  In lower-mass stars, the X-ray 
emission is reminiscent of X-rays observed on the Sun, particularly 
the plasma explosively heated and confined in magnetic loops 
following magnetic reconnection events.  X-ray flares with 
luminosities orders of magnitude more powerful than seen in the 
contemporary Sun are frequently seen in young stars. Evidence for an 
impulsive phase is seen in radio bursts and in U band enhancements 
preceding X-ray flares, thought to be due to the bombardment of the 
stellar surface by electron beams. Thus, young stars prolifically 
accelerate particles to relativistic energies.  In rich young stellar 
clusters, X-rays are also produced by shocks in O star winds, on both 
small ($< 10^2$ R$_\star$) and large (parsec) scales.  If the region 
has been producing rich clusters for a sufficiently long time, the 
resulting supernova remnants will dominate the X-ray properties.  

X-ray studies with the {\it Chandra} and {\it XMM-Newton} space 
observatories are propelling advances of our knowledge and 
understanding of high energy processes during the earliest phases of 
stellar evolution.  In the nearest young stars and clusters ($d < 
500$ pc), they provide detailed information about magnetic 
reconnection processes.  In the more distant and richer regions, the 
X-ray images are amazingly complex with diffuse plasma surrounding 
hundreds of stars exhibiting a wide range of absorptions.  We 
concentrate here on results from three recent large surveys: the {\it 
Chandra Orion Ultradeep Project} (COUP) based on a nearly-continuous 
13-day observation of the Orion Nebula region in 2003, the {\it 
XMM-Newton Extended Survey of Taurus} (XEST) that maps $\sim 
5$~square degrees of the Taurus Molecular Cloud (TMC), and an 
on-going {\it Chandra} survey of high mass star formation regions 
across the Galactic disk.  Because the XEST study is discussed in 
specific detail in a closely  related chapter ({\em G\"udel et al.}, this 
volume) together with optical and infrared surveys, we present only 
selected XEST results. This volume has another closely related 
chapter:  {\em Bally et al.} discuss X-ray emission from high-velocity 
protostellar Herbig-Haro outflows.  The reader interested in earlier 
X-ray studies is referred to reviews by {\em Feigelson and Montmerle} 
(1999), {\em Glassgold et al.} (2000) in Protostars and Planets IV, 
{\em Favata and Micela} (2003), {\em Paerels and Kahn} (2003), and 
{\em G\"udel} (2004). 

The COUP is particularly valuable in establishing a comprehensive 
observational basis for describing the physical characteristics of 
flaring phenomena and elucidating the mechanisms of X-ray production.  
The central portion of the COUP image, showing the PMS population 
around the bright Trapezium stars and the embedded OMC-1 populations, 
is shown in Plate 1 ({\em Getman et al.}, 2005a).  X-rays are 
detected from nearly all known optical members except for many of the 
bolometrically fainter M stars and brown dwarfs. Conversely, 1315 of 
1616 COUP sources (81\%) have clear cluster member counterparts and 
$\simeq$75 (5\%) are new obscured cloud members; most of the 
remaining X-ray sources are extragalactic background sources seen 
through the cloud ({\em Getman et al.}, 2005b).  

X-ray emission and flaring is thus ubiquitous in PMS stars across the 
Initial Mass Function (IMF).  The X-ray luminosity function (XLF) is 
broad, spanning $28 < \log L_x[erg/s] < 32$ ($0.5-8$ keV), with a 
peak around $\log L_x[erg/s] \sim 29$ ({\em Feigelson et al.}, 2005). 
For comparison, the contemporary Sun emits $26 < \log L_x[erg/s] < 
27$, with flares up to $10^{28}$ erg/s, in the same spectral band. 
Results from the more distributed star formation clouds surveyed by 
XEST reveal a very similar X-ray population as in the rich cluster of 
the Orion Nebula, although confined to stars with masses mostly below 
2$M_{\odot}$ (see the chapter by {\em G\"udel et al.}), although there is 
some evidence the XLF is not identical in all regions (Section 4.1). There 
is no evidence for an X-ray-quiet, non-flaring PMS population. 

The empirical findings generate discussion on a variety of 
astrophysical implications including: the nature of magnetic fields 
in young stellar systems; the role of accretion in X-ray emission; 
the effects of X-ray irradiation of protoplanetary disks and 
molecular clouds; and the discovery of X-ray flows from HII regions. 
A number of important related issues are not discussed here 
including: discovery of heavily obscured X-ray populations; X-ray 
identification of older PMS stars; the X-ray emission of 
intermediate-mass Herbig Ae/Be stars; the enigmatic X-ray spectra of 
some O stars; the generation of superbubbles by OB clusters and their 
multiple supernova remnants; and the large-scale starburst conditions 
in the Galactic Center region and other galactic nuclei.  

\bigskip
{\textbf{ 2. FLARING IN PRE-MAIN SEQUENCE STARS}}
\bigskip

\bigskip
\noindent \textbf{2.1 The solar model}
\bigskip

Many lines of evidence link the PMS X-ray properties to magnetic 
activity on the Sun and other late-type magnetically active stars 
such as dMe flare stars, spotted BY Dra variables, and tidally 
spun-up RS CVn post-main sequence binaries.  These systems have 
geometrically complex multipolar magnetic fields in arcades of loops 
rooted in the stellar photospheres and extending into the coronae.  
The field lines become twisted and tangled by gas convection and 
undergo explosive magnetic reconnection.  The reconnection 
immediately accelerates a population of particles with energies tens 
of keV to several MeV; this is the "impulsive phase" manifested by 
gyrosynchrotron radio continuum emission, blue optical/UV continuum 
and, in the Sun, high gamma-ray and energetic particle fluences.  
These particles impact the stellar surface at the magnetic 
footprints, immediately heating gas which flows upward to fill 
coronal loops with X-ray emitting plasma.  It is this "gradual phase" 
of the flare which is seen with X-ray telescopes.  {\em Schrijver 
\& Zwaan} (2000) and {\em Priest \& Forbes} (2002) review the 
observations and physics of solar and stellar flares. 

Extensive multiwavelength properties of PMS stars indicate they are 
highly magnetically active stars ({\em Feigelson and Montmerle}, 
1999).  Hundreds of Orion stars, and many in other young stellar 
populations, have periodic photometric variations from rotationally 
modulated starspots covering $10-50$\% of the surface ({\em Herbst et 
al.}, 2002). A few of these have been subject to detailed Doppler 
mapping showing spot structure.  Radio gyrosynchrotron emission from 
flare electrons spiraling in magnetic loops has been detected in 
several dozen PMS stars ({\em G\"udel}, 2002).  A few nearby PMS 
stars have Zeeman measurements indicating that kilo-Gauss fields 
cover much of the surface ({\em Johns-Krull et al.}, 2004).  In the 
COUP and XEST studies, temperatures inferred from time-averaged 
spectra extends the $T_{\rm cool} - T_{\rm hot}$ and $T - L_x$ trends 
found in the Sun and older stars to higher levels ({\em Preibisch et 
al.}, 2005; {\em Telleschi et al.}, in preparation).  X-ray spectra 
also show plasma abundance anomalies that are virtually identical to 
those seen in older magnetically active stars ({\em Scelsi et al.}, 
2005; {\em Maggio et al.}, in preparation). 

\begin{figure*}
 \epsscale{1.3}
 \plotone{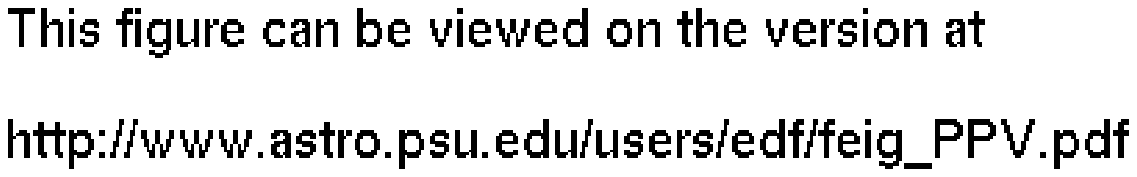}
 \epsscale{1.3}
 \plotone{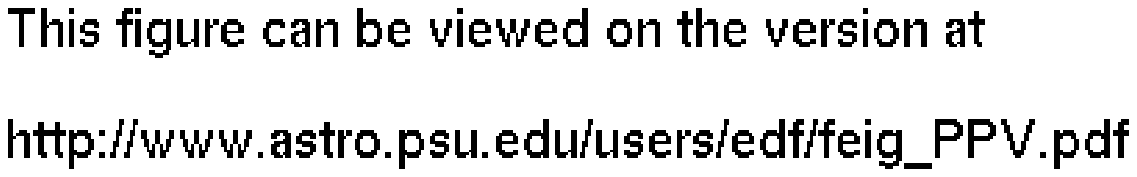}
\caption{\small Two of 1400+ X-ray lightcurves from the {\it Chandra} 
Orion Ultradeep Project.  The abscissa is time spanning 13.2 days, 
and the ordinate gives X-ray count rate in the $0.5-8$ keV band.  (a) 
An Orion star showing typical PMS flaring behavior superposed on a 
rotational modulation of the `characteristic' emission. ({\em 
Flaccomio et al.}, 2005) (b) COUP \#1343 = JW 793, a 
poorly-characterized PMS star in the Orion Nebula, showing a 
spectacular solar-type flare. ({\em Wolk et al.}, 2005; {\em Favata 
et al.}, 2005a) \label{COUP_lc.fig} }
\end{figure*}

Taking advantage of the unprecedented length of the COUP observation, 
{\em Flaccomio et al.}\ (2005) find rotational modulation of X-ray emission 
in at least 10\% of Orion PMS stars with previously determined 
rotation periods from optical monitoring.  An example is shown in 
Fig.\ \ref{COUP_lc.fig}a.  Amplitudes of variability range from 
20--70\% and X-ray periods generally agree with the optical periods. 
In a few cases, it is half the optical value, implying 
X-ray--emitting regions on opposite hemispheres. This result 
indicates that in at least some PMS stars, the X-rays are emitting 
from relatively long-lived structures lying low in the corona that 
are inhomogeneously distributed around the star.  Similar X-ray 
rotational modulations are seen in the Sun and a few older stars. 

{\em Wolk et al.} (2005) examined the flaring behavior of a complete 
sample of solar analogs ($0.9 < M/{\rm M}_\odot < 1.2$) in the Orion 
Nebula Cluster.  The lightcurve in Fig.\ \ref{COUP_lc.fig}b  showing 
one of the more spectacular flares in this subsample reaching $\log 
L_x(peak)[erg/s]=32.2$. Most flares show solar-type rapid rise and 
slower decay; the decay phase can last from $<$1 hour to $>$3 days. 
The brightness and spectral variations during the decay phases of 
this and similarly powerful Orion flares have been analyzed by {\em 
Favata et al.} (2005a) using a loop model previously applied to 
"gradual" (i.e., powerful events spanning several hours) solar and 
stellar flares. The result for COUP \#1343 and other morphologically 
simple cases is clear: the drop in X-ray emission and plasma 
temperature seen in PMS stellar flares is completely compatible with 
that of older stars.  In some COUP flares, the decay shows evidence 
of continued or episodic reheating after the flare peak, a phenomenon 
also seen in solar flares and in older stars.  

The intensity of PMS flaring is remarkably high.  In the solar analog
sample, flares brighter than $L_x(peak) \geq 2 \times 10^{30}$ erg/s
occur roughly once a week ({\em Wolk et al.}, 2005).  The most
powerful flares have peak luminosities up to several times $10^{32}$
erg/s ({\em Favata et al.}, 2005a).  The peak plasma temperature are
typically $T(peak) \simeq 100$ MK but sometimes appear much higher.
The time-integrated energy emitted in the X-ray band during flares in
solar-mass COUP stars is $\log E_x[erg] \simeq 34-36$.  An even more
remarkable flare with $\log E_x[erg] \simeq 37$ was seen by ROSAT from
the non-accreting Orion star Parenago 1724 in 1992 ({\em Preibisch et
al.}, 1995).  These values are far above solar flaring levels:  the
COUP flares are $10^2$ times stronger and $10^2$ times more frequent
than the most powerful flares seen in the contemporary Sun; the
implied fluence of energetic particles may be $10^5$ times above solar
levels ({\em Feigelson et al.}, 2002).

The Orion solar analogs emit a relatively constant `characteristic' 
X-ray level about three-fourths of their time (see Fig.\ 
\ref{COUP_lc.fig}). The X-ray spectrum of this characteristic state 
can be modeled as a two-temperature plasma with one component $T_{\rm 
cool} \simeq 10$ MK and the other component $T_{\rm hot} \simeq 30$ 
MK.  These temperatures are much higher than the quiescent solar 
corona. The concept of ``microflaring'' or ``nanoflaring'' for the 
Sun has been widely discussed ({\em Parker} 1988) and has gained 
favor in studies of older magnetically active stars based on light 
curve and spectral analysis ({\em Kashyap et al.}, 2002; {\em G\"udel 
et al.}, 2003, {\em Arzner \& G\"udel}, 2004).  These latter studies 
of dMe flare stars indicate that a power-law distribution of flare 
energies, $dN/dE \propto E^{-\alpha}$, is present with $\alpha \simeq 
2.0-2.7$.  The energetics is clearly dominated by smaller flares.   
The COUP lightcurves vary widely in appearance, but collectively can 
also be roughly simulated by a powerlaw with $\alpha = 2.0-2.5$ 
without a truly quiescent component ({\em Flaccomio et al.}, 2005). 
Thus, when reference is made to the more easily studied superflares, 
one must always remember that many more weaker flares are present and 
may have comparable or greater astrophysical effects. Not 
infrequently, secondary flares and reheating events are seen 
superposed on the decay phase of powerful flares (e.g., {\em Gagn\'e 
et al.}, 2004; {\em Favata et al.}, 2005a). 

One puzzle with a solar model for PMS flares is that some show 
unusually slow rises.  The non-accreting star LkH$\alpha$ 312 
exhibited a 2-hour fast rise with peak temperature $T \simeq 88$ MK, 
followed by a 6 hour slower rise to $\log L_x(peak)[erg/s] = 32.0$ 
({\em Grosso et al.}, 2004).  The flare from the Class I protostar YLW 
16A in the dense core of the $\rho$ Ophiuchi cloud showed a 
remarkable morphology with two rise phases and similar temperature 
structure (Fig.\ \ref{YLW16a.fig}; {\em Imanishi et al.} 2001).  
Other flares seen with COUP show roughly symmetrical rise and fall 
morphologies, sometimes extending over 1-2 days ({\em Wolk et al.}, 
2005).  It is possible that some of these variations are due to the 
stellar rotation where X-ray structures emerging from eclipse, but 
they are currently poorly understood.  

\begin{figure*}
 \epsscale{0.9} 
 \plotone{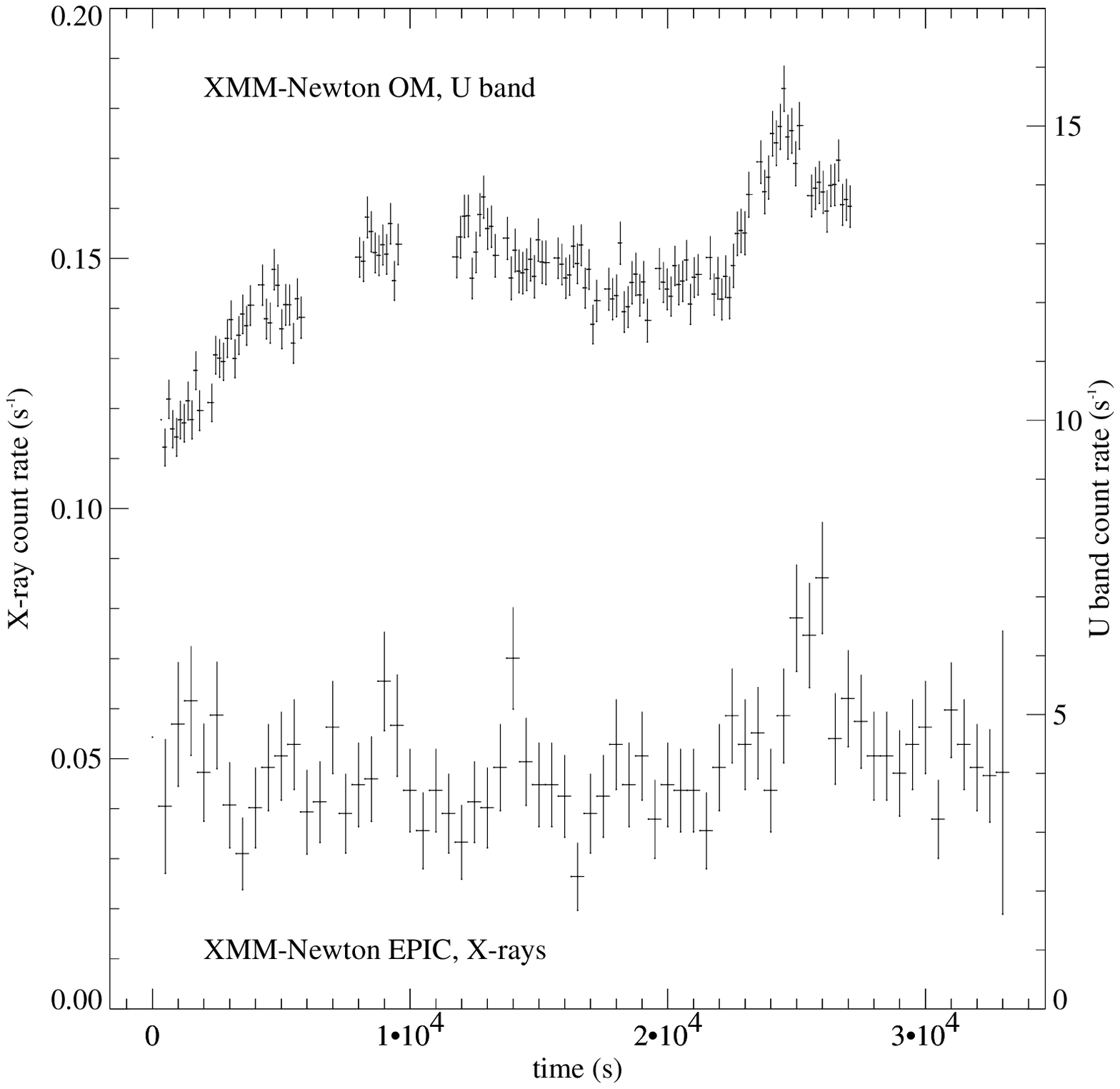} 
 \plotone{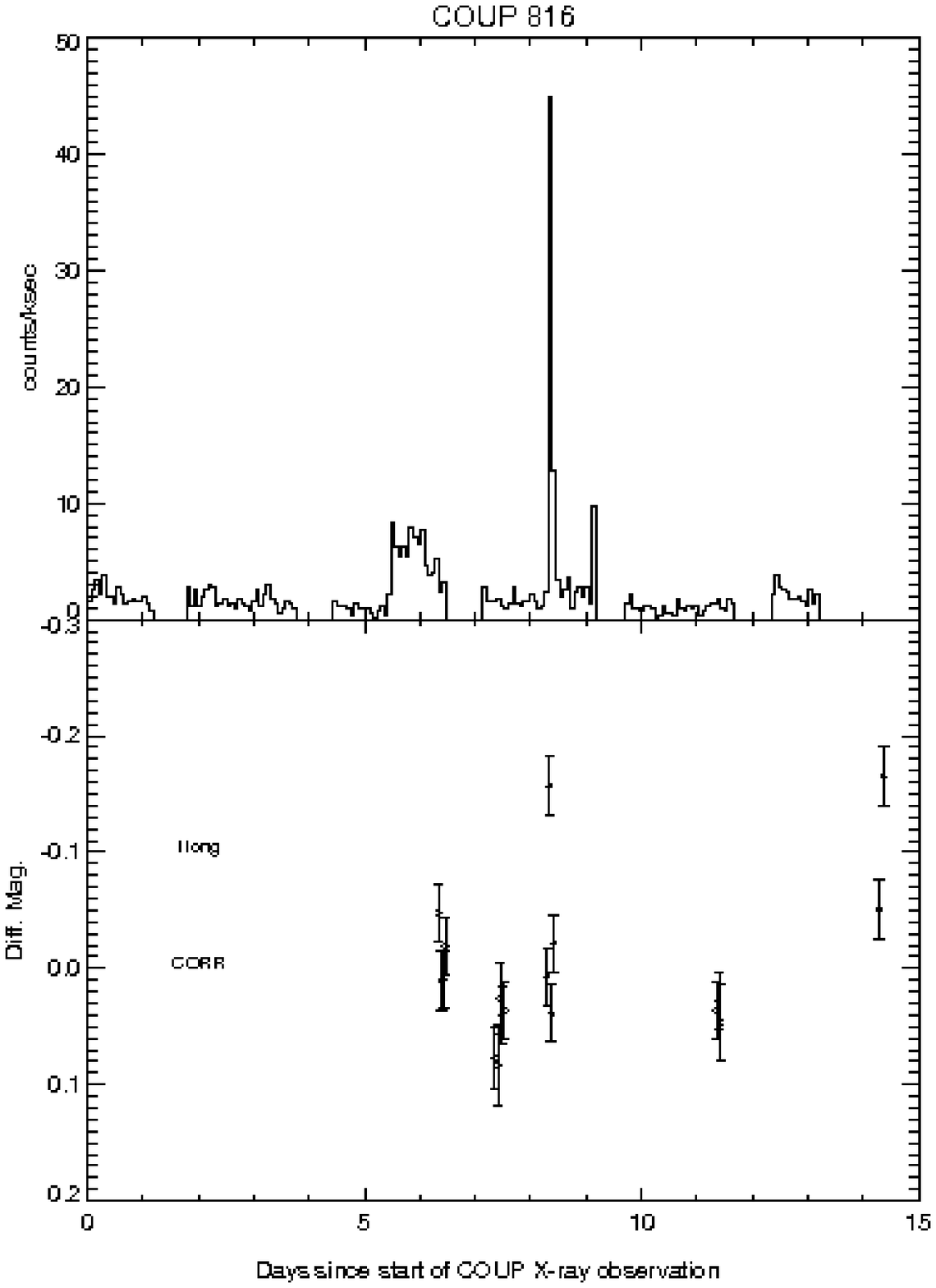}
\caption{\small Detection of the `white light' component during the 
impulsive phases of PMS X-ray flares.  (a) Short-term behavior of the 
classical T Tau binary GK Tau in $U$-band light (upper curve) and 
X-rays (lower curve) from the XEST survey. The light curve covers 
approximately 9 hours. ({\em Audard et al.}, in preparation) (b) 
Rapid X-ray flare (top panel) from COUP \#816 = JW 522, an obscured 
PMS star in the Orion Nebula Cluster, apparently accompanied by 
impulsive $I$-band emission. This COUP X-ray lightcurve spans 13.2 
days. ({\em Stassun et al.}, in preparation) \label{impulsive.fig}}
\end{figure*}

By monitoring young stars with optical telescopes simultaneous with 
X-ray observations, the early impulsive phase of PMS flares can be 
revealed.  This has been achieved with distributed ground-based 
telescopes and in space: the {\it XMM-Newton} satellite has an 
optical-band telescope coaligned with the X-ray telescope. During the 
impulsive phase, electron beams accelerated after the reconnection 
event bombard the stellar chromosphere which produces a burst of 
short-wavelength optical and UV radiation. {\it XMM-Newton} 
observation of the Taurus PMS star GK Tau shows both uncorrelated 
modulations as well as a strong U band burst preceding an X-ray flare 
in good analogy with solar events ({\em Audard et al.}, in 
preparation; Fig.\ \ref{impulsive.fig}a). Ground-based optical and 
H$\alpha$ monitoring of the Orion Nebula during the COUP campaign 
revealed one case of an $I$-band spike simultaneous with a very short 
X-ray flare of intermediate brightness ({\em Stassun et al.}, in 
preparation; Fig.\ \ref{impulsive.fig}b). 

\begin{figure*}
\epsscale{1.8} \plottwo{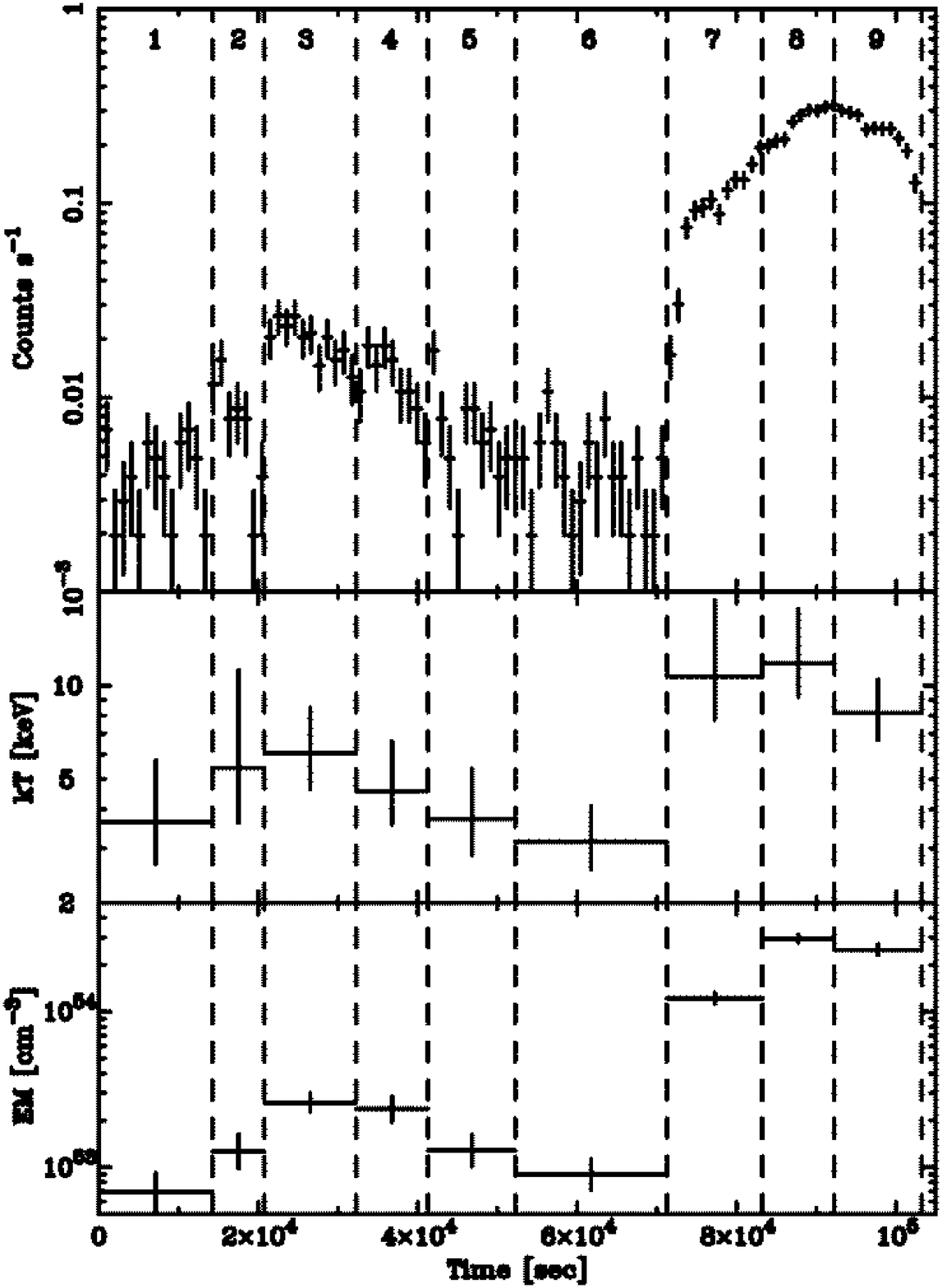}{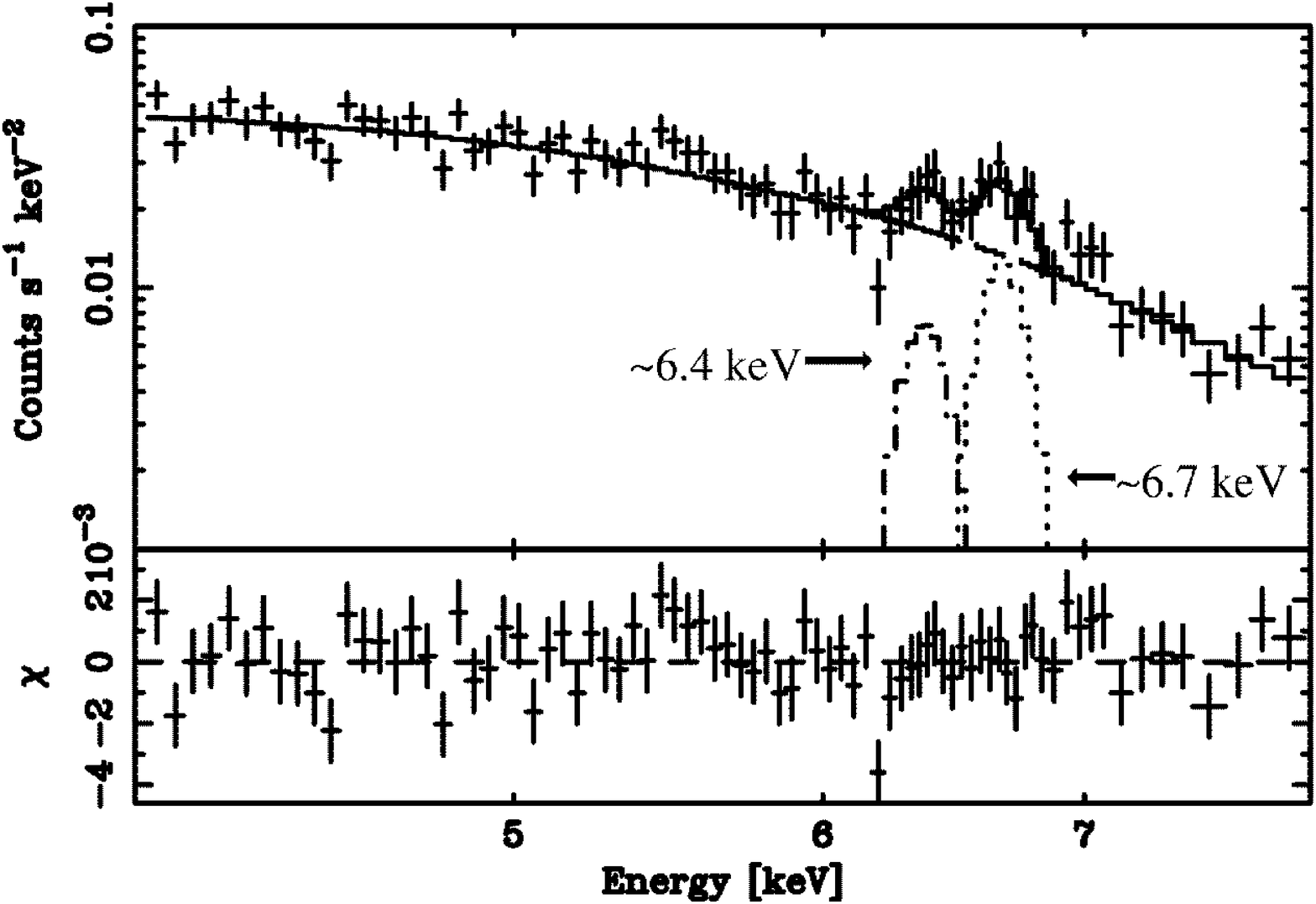} 
\caption{\small X-ray flare from Class I protostar YLW 16A in the 
$\rho$ Ophiuchi cloud, observed with {\it Chandra}.  The flare has an 
unusual morphology and the spectrum shows very hot plasma 
temperatures with strong emission from the fluorescent 6.4 keV line 
of neutral iron attributable to reflection off of the protoplanetary 
disk.  ({\em Imanishi et al.} 2001) \label{YLW16a.fig}}
\end{figure*}

\bigskip
\noindent \textbf{2.2 The role of accretion}
\bigskip

It was established in the 1980-90s that elevated levels of X-ray 
emission in PMS stars appears in both `classical' T Tauri stars, 
with optical/infrared signatures of accretion from a protoplanetary 
disk, and `weak-lined' T Tauri stars without these signatures.  This 
basic result is confirmed but with some important refinements -- and 
controversy -- from recent studies.  

While the presence or absence of a $K$-band emitting inner disk does 
not appear to influence X-ray emission, the presence of accretion has 
a {\it negative} impact on X-ray production ({\em Flaccomio et al.}, 
2003; {\em Stassun et al.}, 2004; {\em Preibisch et al.}, 2005; {\em 
Telleschi et al.}, in preparation).  This is manifested as a 
statistical decrease in X-rays by a factor of $2-3$ in accreting 
$vs.$ non-accreting PMS stars, even after dependencies on mass and 
age are taken into account.  The effect does not appear to arise from 
absorption by accreting gas; e.g., the offset appears in the hard 
$2-8$ keV band where absorption is negligible.  The offset is 
relatively small compared to the $10^4$ range in the PMS X-ray 
luminosity function, and flaring behavior is not affected in any 
obvious way.  One possible explanation is that mass-loaded accreting field 
lines cannot emit X-rays ({\em Preibisch et al.}, 2005). If a 
magnetic reconnection event liberates a certain amount of energy, 
this energy would heat the low-density plasma of non-accretors to 
X-ray emitting temperatures, while the denser plasma in the 
mass-loaded magnetic field lines would be heated to much lower 
temperatures. The remaining field lines which are not linked to the 
disk would have low coronal densities and continue to produce 
solar-like flares.  Note that the very young accreting star XZ Tau 
shows unusual temporal variations in X-ray absorption that can be 
attributed to eclipses by the accretion stream ({\em Favata et al.}, 
2003).  

The optical observations conducted simultaneous with the COUP X-ray 
observations give conclusive evidence that accretion does not produce 
or suppress flaring in the great majority of PMS stars ({\em Stassun 
et al.}, in preparation).  Of the 278 Orion stars exhibiting 
variations in both optical and X-ray bands, not a single case is 
found where optical variations (attributable to either rotationally 
modulated starspots or to changes in accretion) have an evident 
effect on the X-ray flaring or characteristic emission.  

An example from the XEST survey is shown in Fig. 2a where the slow 
modulation seen in the first half is too rapid for effects due to   
rotation, but on the other hand shows no equivalent signatures in 
X-rays.  The optical fluctuations are therefore unrelated to flare 
processes and, in this case, are likely due to variable accretion 
({\em Audard et al.}, in preparation). Similarly, a Taurus brown 
dwarf with no X-ray emission detected in XEST showed a slow rise by a 
factor of six over eight hours in the $U$-band flux ({\em Grosso et 
al.}, in preparation).  Such behavior is uncommon for a flare, and 
because this brown dwarf is accreting, mass streams may again be 
responsible for producing the excess ultraviolet flux.

The simplest interpretation of the absence of statistical links 
between accretion and X-ray luminosities and the absence of
simultaneous optical/X-ray variability is that different magnetic 
field lines are involved with funneling gas from the disk and with 
reconnection events producing X-ray plasma.  There is no evidence 
that the expected shock at the base of the accretion column produces 
the X-rays seen in COUP and XEST. 

There are some counterindications to these conclusions.  A huge 
increase in X-ray emission was seen on long timescales from the star 
illuminating McNeil's Nebula which exhibited an optical/infrared 
outburst attributed to the onset of rapid accretion ({\em Kastner et 
al.}, 2004).  In contrast, however, X-rays are seen before, during 
and after outburst of the EXOR star V1118 Ori, with a lower 
temperature seen when accretion was strongest ({\em Audard et al.}, 
2005).  These findings suggest that accretion, and perhaps the inner 
disk structure, might sometimes affect magnetic field configurations 
and flaring in complicated ways. 

\begin{figure*}
 \epsscale{1.9} 
 \plotone{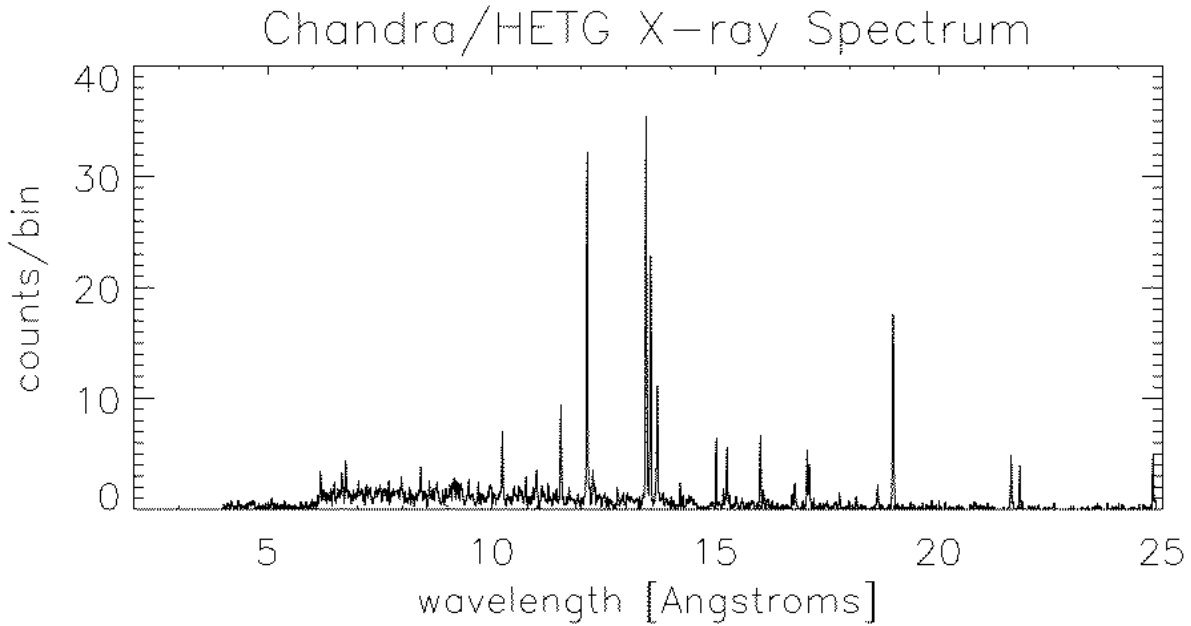} 
\caption{\small High-resolution transmission grating spectrum of the 
nearest classical T Tauri star, TW Hya.  The spectrum is softer than 
other PMS stars, and the triplet line ratios imply either X-ray 
production in a high-density accretion shock or irradiation by 
ultraviolet radiation. ({\em Kastner et al.}, 2002) 
\label{TWHya_spec.fig}}
\end{figure*}

The biggest challenge comes from TW Hya, the nearest and brightest 
accreting PMS star.  It has an X-ray spectrum much softer than most 
COUP or other PMS sources (Fig.\ \ref{TWHya_spec.fig}).  Due to its 
proximity to the Sun, TW Hya is sufficiently bright in X-rays to be 
subject to detailed high-resolution spectroscopy using transmission 
gratings on the {\it Chandra} and {\it XMM-Newton} telescopes ({\em 
Kastner et al.}, 2002; {\em Stelzer and Schmitt}, 2004; {\em Ness and 
Schmitt}, 2005).  According to the magnetospheric accretion scenario, 
accreted material crashes onto the stellar surface with velocities of 
up to several hundred km/sec, which should cause $\sim 10^6$ K shocks 
in which strong optical and UV excess emission and perhaps also soft 
X-ray emission is produced ({\em Lamzin}, 1999).  Density-sensitive 
triplet line ratios of Ne IX and O VII are saturated indicating 
either that the emitting plasma has densities $\log 
n_e$[cm$^{-3}$]$\sim 13$, considerably higher than $\log 
n_e$[cm$^{-3}$]$\sim 10$ characteristic of low-level coronal emission 
although reminiscent of densities in flares. However, these densities 
were measured during an observation dominated by relative quiescence 
with no hot plasma present. Alternatively, the high triplet ratios 
might be induced by plasma subject to strong ultraviolet irradiation.  
A similar weak forbidden line in the O VII triplet is seen in the 
accreting PMS star BP Tau ({\em Schmitt et al.}, 2005), and similar
soft X-ray emission is seen from the Herbig Ae star HD 163296
({\em Swartz et al.}, 2005).  

If the plasma material in TW Hya is drawn from the disk rather than 
the stellar surface, one must explain the observed high Ne/Fe 
abundance ratio which is similar to that seen in flare plasmas.  One 
possibility is that the abundance anomalies do not arise from the 
coronal First Ionization Potential effect, but rather from the 
depletion of refractory elements into disk solids ({\em Brinkman et 
al.}, 2001; {\em Drake et al.}, 2005).  This model, however, must 
confront models of the infrared disk indicating that grains have 
sublimated in the disk around 4 AU, returning refractory elements 
back into the gas phase ({\em Calvet et al.}, 2002).   

Finally, we note that current X-ray instrumentation used for PMS 
imaging studies is not very sensitive to the cooler plasma expected 
from accretion shocks, and that much of this emission may be 
attenuated by line-of-sight interstellar material.  The possibility 
that some soft accretion X-ray emission is present in addition to the 
hard flare emission is difficult to firmly exclude.  But there is 
little doubt that most of the X-rays seen with {\it Chandra} and {\it 
XMM-Newton} are generated by magnetic reconnection flaring rather 
than the accretion process. 

\bigskip
\noindent \textbf{2.3 The role of disks}
\bigskip

There are strong reasons from theoretical models to believe that PMS 
stars are magnetically coupled to their disks at the corotation radii 
typically $5-10$ R$_\star$ from the stellar surface (e.g., {{\em 
Hartmann} 1998; \em Shu et al.}, 2000).  This hypothesis unifies such 
diverse phenomena as the self-absorbed optical emission lines, the 
slow rotation of accreting PMS stars, and the magneto-centrifugal 
acceleration of Herbig-Haro jets. However, there is little {\it 
direct} evidence for magnetic field lines connecting the star and the 
disk. Direct imaging of large-scale magnetic fields in PMS stars is 
only possible today using Very Long Baseline Interferometry at radio 
wavelengths where an angular resolution of 1 mas corresponds to 
0.14~AU at the distance of the Taurus or Ophiuchus clouds. But only a 
few PMS stars are sufficiently bright in radio continuum for such 
study. One of the components of T Tau S has consistently shown 
evidence of magnetic field extensions to several stellar radii, 
perhaps connecting to the inner border of the accretion disk ({\em 
Loinard et al.} 2005). 

But X-ray flares can provide supporting evidence for star-disk 
magnetic coupling.  An early report of star-disk fields arose from a 
sequence of three powerful flares with separations of $\sim 20$ hr 
from the Class~I protostar YLW~15 in the $\rho$~Oph cloud ({\em 
Tsuboi et al.}, 2000).  Standard flare plasma models indicated loop 
lengths around 14 R$_\odot$, and periodicity might arise from 
incomplete rotational star-disk coupling ({\em Montmerle et al.}, 
2000).  However, it is also possible that the YLA~15 flaring is not 
truly periodic; many cases of multiple flares without periodicities 
are seen in the COUP lightcurves.  

Analysis of the most luminous X-ray flares in the COUP study also 
indicates that huge magnetic structures can be present. {\em Favata 
et al.} (2005a) reports analysis of the flare decay phases in sources 
such as COUP \#1343 (Fig.\ \ref{COUP_lc.fig}) using models that 
account for reheating processes, which otherwise can lead to 
overestimation of loop lengths.  The combination of very high 
luminosities ($\log L_x(peak)[erg/s] \simeq 31-32$), peak 
temperatures in excess of 100~MK, and very slow decays appears to 
require loops much larger than the star, up to several $10^{12}$ cm 
or 5--20 R$_\star$. Recall that these flares represent only the 
strongest $\sim 1$\% of all flares observed by COUP; most flares from 
PMS stars are much weaker and likely arise from smaller loops.  This 
is clearly shown in some stars by the rotational modulation of the 
non-flaring component in the COUP study ({\em Flaccomio et al.}, 
2005).  

Given the typical $2-10$ day rotation periods of PMS stars, is seems 
very doubtful such long flaring loops would be stable if both 
footpoints were anchored to the photosphere.  Even if MHD 
instabilities are not important, gas pressure and centrifugal forces 
from the embedded plasma may be sufficient to destroy such enormous 
coronal loops ({\em Jardine and Unruh}, 1999). {\em Jardine et al.} 
(2006) develop a model of magnetically-confined multipolar coronae of 
PMS stars where accretion follows some field lines while others 
contain X-ray emitting plasma; the model also accounts for observed 
statistical relations between X-ray properties and stellar mass.  

The magnetospheres of PMS stars are thus likely to be quite complex. 
Unlike the Sun where only a tiny fraction of the photosphere has 
active regions, intense multipolar fields cover much of the surface 
in extremely young stars.  Continuous microflaring is likely 
responsible for the ubiquitous strong $10-30$ MK plasma emission.  
Other field lines extend several stellar radii: some are mass-loaded 
with gas accreting from the circumstellar disk, while others may 
undergo reconnection producing the most X-ray-luminous flares.  

\begin{figure*}
 \epsscale{2.2} 
 \includegraphics[scale=0.6]{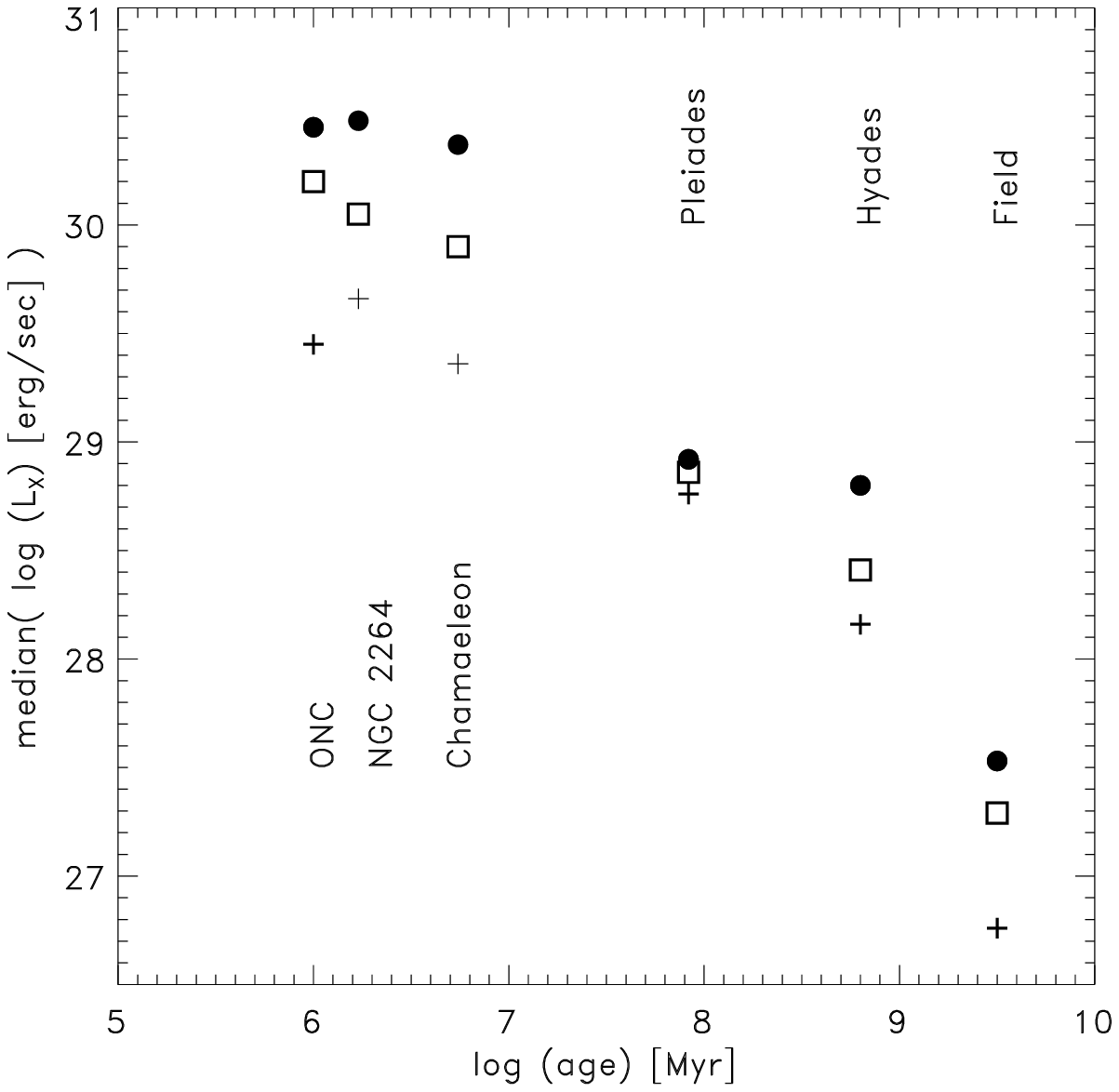}
 \includegraphics[scale=0.8,angle=90.]{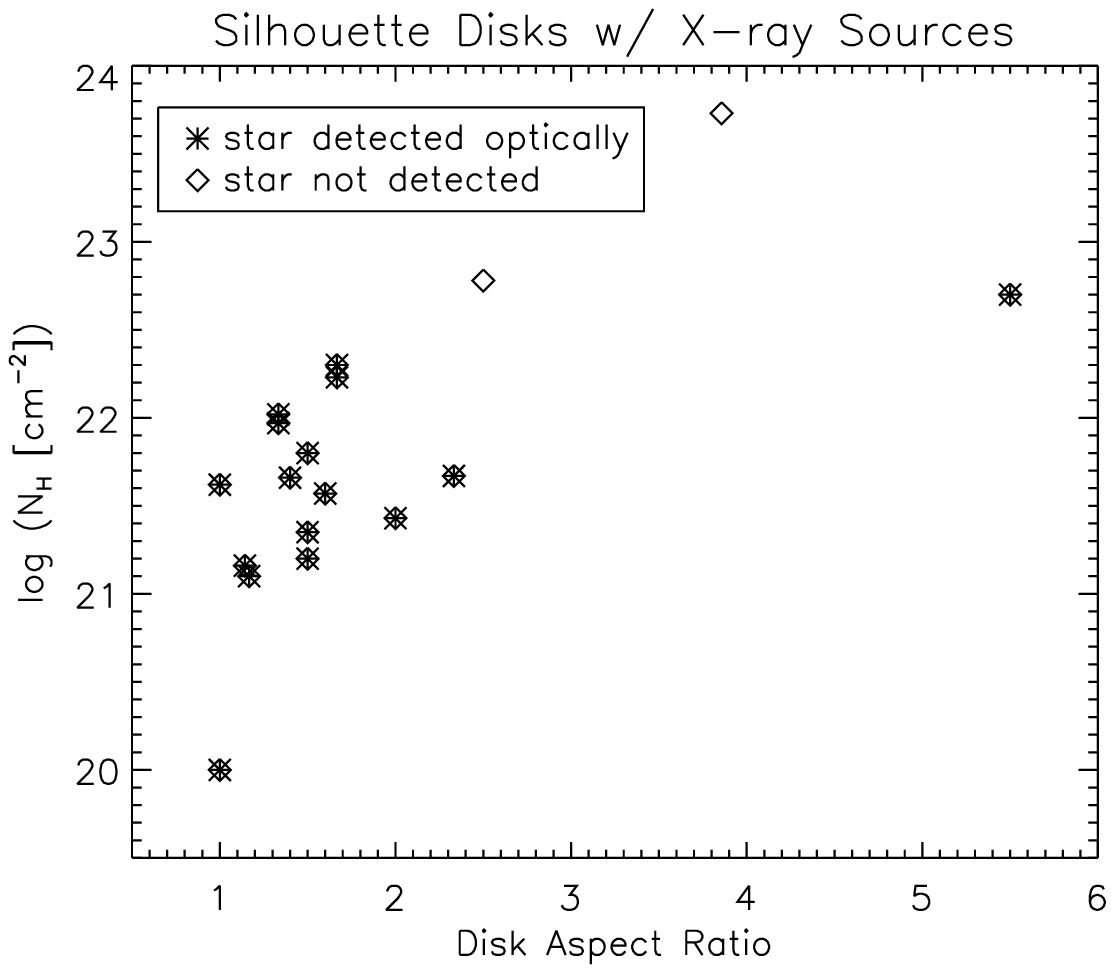} 
 \caption{\small (a)  Evolution of the median X-ray luminosities 
for stars in different mass ranges: $0.9-1.2$ M$_\odot$ (solid 
circles), $0.5-0.9$ M$_\odot$ (open squares), and $0.1-0.5$ M$_\odot$ 
(plusses). ({\em Preibisch and Feigelson}, 2005)  (b) The link between 
soft X-ray absorption and proplyd inclination is the first 
measurement of gas column densities in irradiated disks. ({\em 
Kastner et al.}, 2005) \label{Evol_propl.fig} }
\end{figure*}

\bigskip
{\textbf{ 3. THE EVOLUTION OF MAGNETIC ACTIVITY}}
\bigskip

The COUP observation provides the most sensitive, uniform and 
complete study of X-ray properties for a PMS stellar population 
available to date.  When combined with studies of older stellar 
clusters, such as the Pleiades and Hyades, and of volume-limited 
samples in the solar neighborhood, evolutionary trends in X-ray 
emission can be traced.  Since chromospheric indicators of magnetic 
activity (such as Ca II line emission) are confused by accretion, and 
photospheric variations from rotationally modulated starspots are too 
faint to be generally measured in most older stars, X-ray emission is 
the only magnetic indicator which can be traced in stellar 
populations from $10^5$ to $10^{10}$ yr.  The result from the PMS to 
the gigayear-old disk population is shown in Fig.\ 
\ref{Evol_propl.fig}a ({\em Preibisch and Feigelson}, 2005).  The two 
critical advantages here over other measures of magnetic activity 
evolution are the complete samples (and correct treatment of 
nondetections) and stratification by mass.  The latter is important 
because the mass-dependence of X-ray luminosities (for unknown 
reasons) differs in PMS and main sequence stars ({\em Preibisch et 
al.}, 2005).  

If one approximates the decay of magnetic activity as a powerlaw, 
then evolution in the $0.5 < M < 1.2$ M$_\odot$ mass range is 
approximately power-law with $L_x \propto \tau^{-3/4}$ over the wide 
range of ages $5 < \log \tau[yr] < 9.5$.  Other X-ray studies of 
older disk stars suggest that the decay rate steepens: $L_x \propto 
\tau^{-3/2}$ or $\tau^{-2}$ over $8 < \log\tau[yr] < 10$ ({\em 
G\"udel et al.}, 1997; {\em Feigelson et al.}, 2004). Note, however, 
that {\em Pace and Pasquini} (2004) find no decay in chromospheric 
activity in a sample of solar mass stars after 3 Gyr. These results 
are similar to, but show more rapid decay than, the classical {\em 
Skumanich} (1972) $\tau^{-1/2}$ relation which had been measured for 
main sequence stars only over the limited age range $7.5 < \log 
\tau[yr] < 9.5$.  The COUP sample also exhibits a mild decay in 
magnetic activity for ages $5 < \log\tau[yr] < 7$ within the PMS 
phase, although the trend is dominated by star-to-star scatter ({\em 
Preibisch and Feigelson} 2005).

While these results would appear to confirm and elaborate the 
long-standing rotation-age-activity relationship of solar-type stars, 
the data paint a more complex picture. The {\it Chandra} Orion 
studies show that the rotation-activity relation is largely absent at 
1 Myr ({\em Stassun et al.}, 2004; {\em Preibisch et al.}, 2005). 
This finding suggests the somewhat surprising result that the 
activity-age decay is strong across the entire history of solar-type 
stars but is not entirely attributable to rotational deceleration. 
The PMS magnetic fields may either be generated by a solar-type 
dynamo that is completely saturated, or by a qualitatively different 
dynamo powered by turbulence distributed throughout the convective 
interior rather than by rotational shear at the tachocline. At the 
same time, the Orion studies show a small {\it positive} correlation 
between rotation period and X-ray activity, similar to that seen in 
the ``super-saturated" regime of main sequence stars. It is also 
possible that this effect is due to a sample bias against slowly 
rotating, X-ray-weak Orion stars ({\em Stassun et al.}, 2004). 

The XEST findings in the Taurus PMS population give a different 
result suggesting that an unsaturated solar-type dynamo may in fact 
be present in PMS stars when rotation periods are longer than a few 
days ({\em Briggs et al.}, 2006, in preparation). It is possible that 
the somewhat more evolved Taurus sample, compared to the Orion Nebula 
Cluster stars, has produced sufficiently prominent radiative zones in 
some of these late-type PMS stars to put a solar-type dynamo into 
operation.

The origins of magnetic fields in PMS stars are thus still not well 
established.  It is possible that both tachoclinal and convective 
dynamos are involved, as discussed by {\em Barnes} (2003a,b).  There 
is a hint of a transition between convective and rotational dynamos 
in the plot of $L_x/L_{bol}$ against mass in Orion stars.  The X-ray 
emissivity for many stars drops precipitously for masses above $2-3$ 
M$_\odot$, which is also the boundary between lower-mass convective 
and higher-mass radiative interiors ({\em Feigelson et al.}, 2003). 
Another possible influence is that accretion in younger PMS stars 
alters convection and thereby influences the magnetic field 
generation process ({\em Siess et al.}, 1999; {\em Stassun et al.}, 
2004). 

The magnetic activity history for M stars with masses 0.1 to 0.4 
M$_\odot$ appears to be different than more massive PMS stars (Fig.\ 
\ref{Evol_propl.fig}a). Only a mild decrease in X-ray luminosity, and 
even a mild increase in $L_x/L_{\rm bol}$ and $F_X$, is seen over the 
$6< \log\tau[yr] <8$ range, though the X-ray emission does decay over 
gigayear timescales. This result may be related to the 
well-established fact that the low-mass M stars have much longer 
rotational slow-down times than solar-type stars.  But the difference 
in behavior compared to higher mass stars could support the idea that 
the dynamos in PMS and dM stars both arise from a convective 
turbulent dynamo. These issues are further discussed in {\em Mullan 
and MacDonald} (2001), {\em Feigelson et al.} (2003), {\em Barnes} 
(2003a), and {\em Preibisch et al.} (2005).

An unresolved debate has emerged concerning the onset of X-ray 
emission in PMS stars.  There is no question that magnetic activity 
with violent flaring is common among Class I protostars with ages 
$\sim 10^5$ yr ({\em Imanishi et al.}, 2001; {\em Preibisch}, 2004; 
Fig.\ \ref{YLW16a.fig}).  The question is whether X-ray emission is 
present in Class 0 protostars with ages $\sim 10^4$ yr.  There is one 
report of hard X-rays from two Class 0 protostars in the OMC 2/3 
region ({\em Tsuboi et al.}, 2001); however, other researchers 
classify the systems as Class 1 ({\em Nielbock et al.}, 2003).  An 
X-ray emitting protostar deeply embedded in the R Corona Australis 
cloud core has a similarly uncertain Class 0 or I status ({\em 
Hamaguchi et al.}, 2005).  In contrast, a considerable number of 
well-established Class 0 protostars appear in {\it Chandra} images 
and are not detected, for example, in the NGC~1333 and Serpens 
embedded clusters ({\em Getman et al.}, 2002; {\em Preibisch}, 2004). 
However, because Class 0 stars are typically surrounded by very dense 
gaseous envelopes, it is possible that the X-ray nondetections arise 
from absorption rather than an absence of emission.  An interesting 
new case is an intermediate-mass Class 0 system in the IC 1396N 
region which exhibits extremely strong soft X-ray absorption ({\em 
Getman et al.}, in preparation). 

\bigskip
{\textbf{ 4. X-RAY STARS AND HOT GAS IN \\ MASSIVE STAR-FORMING 
REGIONS}}
\bigskip

Most stars are born in massive star-forming regions (MSFRs) where 
rich clusters containing thousands of stars are produced in molecular 
cloud cores. Yet, surprisingly little is known about the lower mass 
populations of these rich clusters.  Beyond the Orion Molecular 
Clouds, near-infrared surveys like 2MASS are dominated by foreground 
or background stars, and the Initial Mass Functions are typically 
measured statistically rather than by identification of individual 
cluster members.  X-ray surveys of MSFRs are important in this 
respect because they readily discriminate young stars from unrelated 
objects that often contaminate $JHK$ images of such fields, 
especially for those young stars no longer surrounded by a dusty 
circumstellar disk.  Furthermore, modern X-ray telescopes penetrate 
heavy obscuration (routinely $A_V \sim 10-100$~mag, occasionally up 
to 1000~mag) with little source confusion or contamination from 
unrelated objects to reveal the young stellar populations in MSFRs.  

The O and Wolf-Rayet (WR) members of MSFRs have been catalogued, and 
the extent of their UV ionization is known through HII region 
studies. But often little is known about the fate of their powerful 
winds.  The kinetic power of a massive O-star's winds injected into 
its stellar neighborhood over its lifetime is comparable to the input 
of its supernova explosion. Theorists calculate that wind-blown 
bubbles of coronal-temperature gas should be present, but no clear 
measurement of this diffuse plasma had been made in HII regions prior 
to {\it Chandra}.  X-ray studies also detect the presence of earlier 
generations of OB stars through the shocks of their supernova 
remnants (SNRs).  In very rich and long-lived star forming cloud 
complexes, SNRs combine with massive stellar winds to form 
superbubbles and chimneys extending over hundreds of parsecs and 
often into the Galactic halo. O stars are thus the principal drivers 
of the interstellar medium. 

The MSFR X-ray investigations discussed here represent only a 
fraction of this rapidly growing field.  A dozen early observations 
of MSFRs by {\it Chandra} and {\it XMM-Newton} are summarized by {\em 
Townsley et al.} (2003).  Since then, {\it Chandra} has performed 
observations of many other regions, typically revealing hundreds of 
low-mass PMS stars, known and new high-mass OB stars, and 
occasionally diffuse X-ray emission from stellar winds or SNRs. In 
addition to those discussed below, these include NGC 2024 in the 
Orion B molecular cloud ({\em Skinner et al.}, 2003), NGC 6193 in Ara 
OB1 ({\em Skinner et al.}, 2005), NGC 6334 ({\em Ezoe et al.}, 2006), 
NGC 6530 ionizing Messier 8 ({\em Damiani et al.}, 2004), the Arches 
and Quintuplet Galactic Center clusters ({\em Law and Yusef-Zadeh}, 
2004; {\em Rockefeller et al.}, 2005), and Westerlund 1 which has an 
X-ray pulsar ({\em Muno et al.}, 2006; {\em Skinner et al.}, 2006).  
{\it Chandra} studies of NGC 6357, M 16, RCW 49, W 51A, W 3 and other 
regions are also underway.  Both {\it XMM-Newton} and {\it Chandra} 
have examined rich clusters in the Carina Nebula ({\em Evans et al.}, 
2003, 2004; {\em Albacete Colombo et al.}, 2003), NGC 6231 at the 
core of the Sco OB1 association, and portions of Cyg OB2.

\bigskip
\noindent \textbf{4.1 Cepheus B, RCW 38, and stellar populations}
\bigskip

Each {\it Chandra} image of a MSFR shows hundreds, sometimes over a 
thousand, unresolved sources.  For regions at distances around $d 
\simeq 1-3$ kpc, only a small fraction (typically $3-10$\%) of these 
sources are background quasars or field Galactic stars.  The stellar 
contamination is low because PMS stars are typically 100-fold more 
X-ray luminous than $1-10$ Gyr old main sequence stars (Fig.\ 
\ref{Evol_propl.fig}a).  Since {\it Chandra} source positions are 
accurate to $0.2^{\prime\prime}-0.4^{\prime\prime}$, identifications 
have little ambiguity except for components of multiple systems.  The 
XLF of a stellar population spans 4 orders of magnitude; 2 orders of 
magnitude of this range arises from a correlation with stellar mass 
and bolometric magnitude ({\em Preibisch et al.}, 2005).  This means 
that the X-ray flux limit of a MSFR observation roughly has a 
corresponding limit in $K$-band magnitude and mass. Day-long 
exposures of regions $d\simeq 2$ kpc away are typically complete to 
$\log L_x[erg/s] \sim 29.5$ which gives nearly complete samples down 
to $M \simeq 1$ M$_\odot$ with little contamination.  We outline two 
recent studies of this type.

A 27 hr {\it Chandra} exposure of the stellar cluster illuminating 
the HII region RCW 38 ($d \simeq 1.7$ kpc) reveals 461 X-ray sources, 
of which 360 are confirmed cluster members ({\em Wolk et al.}, in 
preparation).  Half have near-infrared counterparts of which 20\% 
have $K$-band excesses associated with optically thick disks.  The 
cluster is centrally concentrated with a half-width of 0.2 pc and a 
central density of 100 X-ray stars/pc$^2$.  Obscuration  of the 
cluster members, seen both in the soft X-ray absorption column and 
near-infrared photometry, is typically $10 < A_V < 20$ mag.  The 
X-ray stars are mostly unstudied; particular noteworthy are 31 X-ray 
stars which may be new obscured OB stars.  Assuming a standard IMF, 
the total cluster membership is estimated to exceed 2000 stars.  
About 15\% of the X-ray sources are variable, and several show plasma 
temperatures exceeding 100 MK.  

A recent {\it Chandra} study was made of the Sharpless 155 HII region 
on the interface where stars from the Cep OB3b association ($d = 725$ 
pc) illuminates the Cepheus B molecular cloud core ({\em Getman et 
al.}, 2006).  Earlier, a few ultracompact HII regions inside the 
cloud indicated an embedded cluster is present, but little was known 
about the embedded population.  The 8 hr exposure shows 431 X-ray 
sources of which 89\% are identified with $K$-band stars.  Sixty-four 
highly-absorbed X-ray stars inside the cloud provide the best census 
of the embedded cluster, while the 321 X-ray stars outside the cloud 
provide the best census of this portion of the Cep OB3b cluster.  
Surprisingly, the XLF of the unobscured sample has a different shape 
from that seen in the Orion Nebula Cluster, with an excess of stars 
around $\log L_x[erg/s] \simeq 29.7$ or $M \simeq 0.3$ M$_\odot$.  It 
is not clear whether this arises from a deviation in the IMF or some 
other cause, such as sequential star formation generating a 
non-coeval population. The diffuse X-rays in this region, which has 
only one known O star, are entirely attributable to the integrated 
contribution of fainter PMS stars.  

\bigskip
\noindent \textbf{4.2 M~17 and X-ray flows in HII regions}
\bigskip

For OB stars excavating an HII region within their nascent molecular
cloud, diffuse X-rays may be generated as fast winds shock the
surrounding media ({\em Weaver et al.}, 1977).  {\it Chandra} has 
clearly discriminated this diffuse emission from the hundreds of 
X-ray-emitting young stars in M~17 and the Rosette Nebula ({\em 
Townsley et al.}, 2003).  

Perhaps the clearest example of diffuse X-ray emission in MSFRs is 
the {\it Chandra} observation of M17, a bright blown-out blister HII 
region on the edge of a massive molecular cloud ($d \simeq 1.6$ kpc). 
The expansion of the blister HII region is triggering star formation 
in its associated giant molecular cloud which contains an 
ultracompact HII region, water masers, and the dense core M17SW.  M17 
has 100 stars earlier than B9 (for comparison, the Orion Nebula 
Cluster has 8), with 14 O stars.  The {\it Chandra} image is shown in 
Plate 3, along with an earlier, wider-field image from the {\it 
ROSAT} satellite. Over 900 point sources in the $\sim 17^{\prime} 
\times 17^{\prime}$ field are found ({\em Broos et al.}, in 
preparation).  

The diffuse emission of M~17 is spatially concentrated eastward of 
the stellar cluster and fills the region delineated by the 
photodissociation region and the molecular cloud.  The X-ray spectrum 
can be modeled as a two-temperature plasma with $T=1.5$~MK and 7~MK, 
and a total intrinsic X-ray luminosity (corrected for absorption) of 
$L_{{\rm x,diffuse}}=3 \times 10^{33}$~erg/s ({\em Townsley et al.}, 
2003).  The X-ray plasma has mass $M \sim 0.1$~M$_\odot$ and density 
0.1--0.3~cm$^{-3}$ spread over several cubic parsecs. It represents 
only $\sim 10^4$ yr of recent O wind production; past wind material 
has already flowed eastward into the Galactic interstellar medium.

The diffuse emission produced by the M~17 cluster, and similar but 
less dramatic emission by the Rosette Nebula cluster, gives new 
insight into HII region physics.  The traditional HII region model 
developed decades ago by Str{\" o}mgren and others omitted the role 
of OB winds which were not discovered until the 1960s.  The winds 
play a small role in the overall energetics of HII regions, but they 
dominate the momentum and dynamics of the nebula with $\frac{1}{2} 
\dot{M} v_w^2 \sim 10^{36-37}$~erg/s for a typical early-O star.  If 
completely surrounded by a cold cloud medium, an O star should create 
a ``wind-swept bubble'' with concentric zones:  a freely expanding 
wind, a wind termination shock followed by an X-ray emitting zone, 
the standard $T=10^4$~K HII region, the ionization front, and the 
interface with the cold interstellar environment ({\em Weaver et 
al.}, 1977; {\em Capriotti and Kozminski}, 2001). 

These early models predicted $L_x \sim 10^{35}$ erg/s from a single 
embedded O star, two orders of magnitude brighter than the emission 
produced by M~17 ({\em Dunne et al.}, 2003).  Several explanations 
for this discrepancy can be envisioned: perhaps the wind energy is 
dissipated in a turbulent mixing layer ({\em Kahn and Breitschwerdt}, 
1990), or the wind terminal shock may be weakened by mass-loading of 
interstellar material (e.g., {\em Pittard et al.}, 2001). Winds from 
several OB stars may collide and shock before they hit the ambient 
medium ({\em Canto et al.}, 2000).  Finally, a simple explanation may 
be that most of the kinetic energy of the O star winds remains in a 
bulk kinetic flow into the Galactic interstellar medium ({\em 
Townsley et al.}, 2003).

\bigskip
\noindent \textbf{4.3 Trumpler 14 in the Carina Nebula}
\bigskip

The Carina complex at $d \simeq 2.8$~kpc, is a remarkably rich 
star-forming region containing 8 open clusters with at least 64 O 
stars, several WR stars, and the luminous blue variable $\eta$ Car.  
The presence of WR stars may indicate past supernovae, although no 
well-defined remnant has been identified.  One of these clusters is 
Tr~14, an extremely rich, young ($\sim$1~My), compact OB association 
with $\sim$30 previously identified OB stars. Together with the 
nearby Trumpler~16 cluster, it has the highest concentration of O3 
stars known in the Galaxy.  Over 20 years ago, an {\em Einstein 
Observatory} X-ray study of the Carina star-forming complex detected 
a few dozen high-mass stars and diffuse emission attributed to O star 
winds ({\em Seward and Chlebowski}, 1982).  {\it Chandra} studies 
show that thousands of the lower-mass stars in these young clusters 
were likely to be contributing to this diffuse flux; a major goal is 
to determine the relative contributions of stars, winds and SNRs to 
the extended emission in the Carina Nebula.  

Plate 4a shows a 16 hr {\it Chandra} exposure centered on HD~93129AB, 
the O2I/O3.5V binary at the center of Tr~14 ({\em Townsley et al.}, 
in preparation).  Over 1600 members of the Tr~14 and Tr~16 clusters 
can be identified from the X-ray point sources and extensive diffuse 
emission is clearly present.  The diffuse emission surrounding Tr~14 
is quite soft with subsolar elemental abundances, similar to the M~17 
OB wind shocks.  But the much brighter diffuse emission seen far from 
the massive stellar clusters, is less absorbed and requires enhanced 
abundances of Ne and Fe. This supports models involving old 
``cavity'' supernova remnants that exploded inside the Carina 
superbubble (e.g.\ {\em Chu et al.}, 1993).

\begin{figure*}[ht]
 \epsscale{2.2} 
 \plottwo{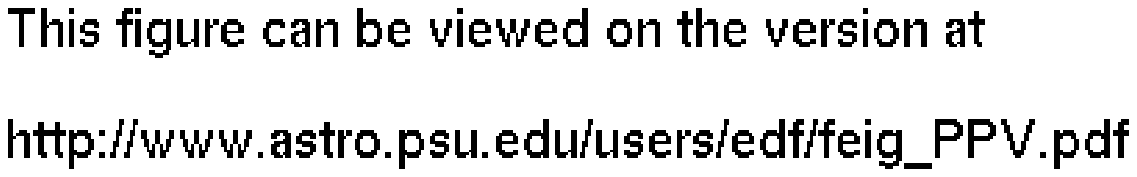}{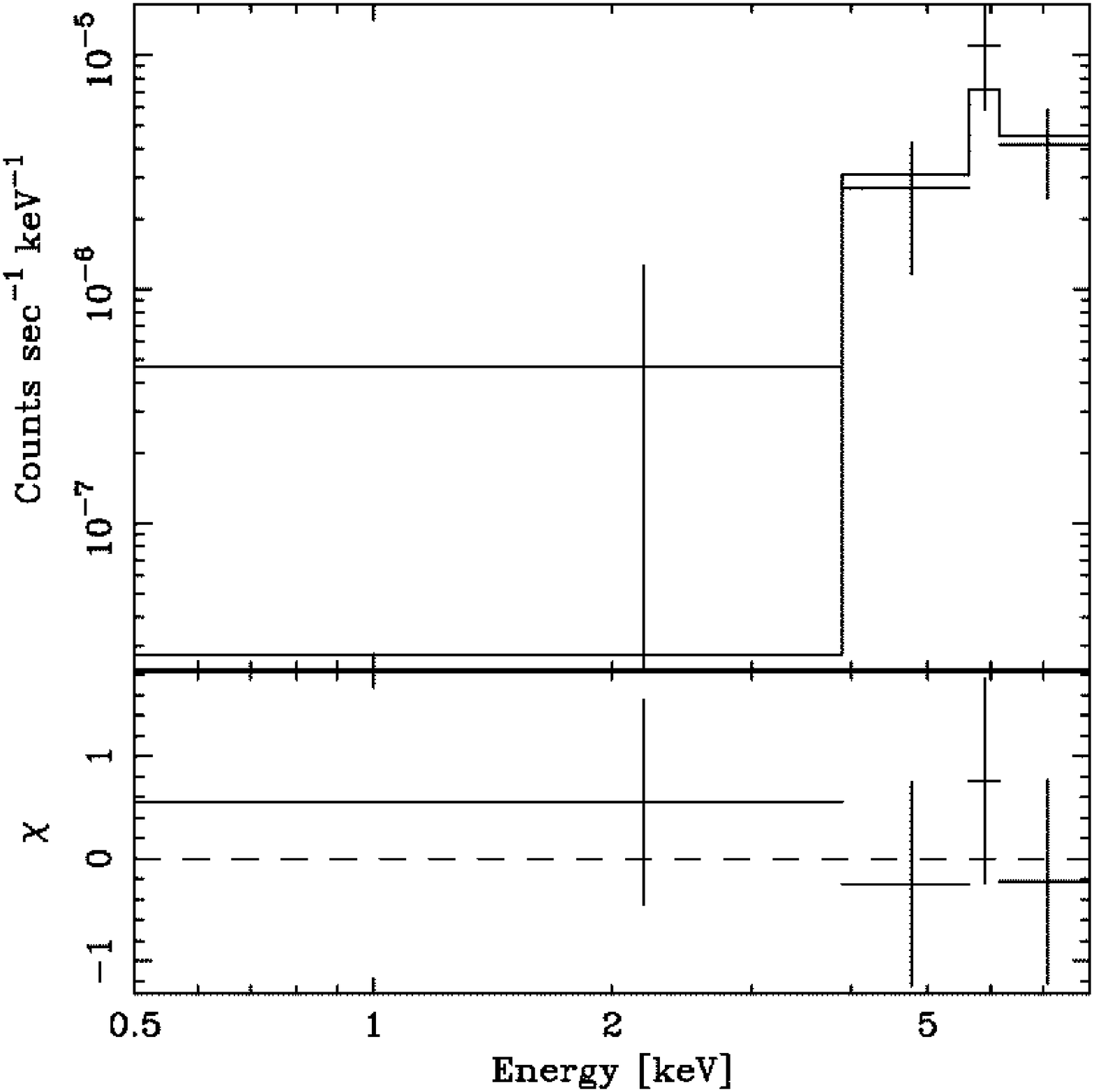} 
 \caption{\small  X-ray properties of two stars deeply embedded
in the OMC-1 South molecular cloud core from the {\it Chandra}
Orion Ultradeep Project. (Left) Lightcurve of COUP 554 over 13.2 days
where the histogram shows the integrated brightness (left-had
vertical axis) and the dots show the energies of individual photons 
(right-hand vertical axis).  (Right) Spectrum of COUP 632 showing 
very strong absorption at energies below 4 keV.   ({\em Grosso et 
al.}, 2005) \label{protostar.fig}}
\end{figure*}

{\it Chandra} resolves the two components of HD~93129AB separated by 
$\sim 3^{\prime\prime}$: HD~93129B shows a typical O-star soft X-ray 
spectrum ($T \sim 6$~MK), while HD~93129A shows a similar soft 
component plus a $T \sim 35$~MK component which dominates the total 
X-ray luminosity. This hard spectrum and high X-ray luminosity are 
indicative of a colliding-wind binary ({\em Pittard et al.}, 2001), 
in agreement with the recent finding that HD~93129A is itself a 
binary ({\em Nelan et al.}, 2004).  Other colliding wind binaries are 
similarly identified in the cluster.

\bigskip
\noindent \textbf{4.4 The starburst of 30 Doradus}
\bigskip

Plate 4b shows a 6 hr {\it Chandra} exposure of 30 Dor in the Large 
Magellanic Cloud, the most luminous Giant Extragalactic HII Region 
and ``starburst cluster'' in the Local Group.  30~Dor is the product 
of multiple epochs of star formation, which have produced multiple 
SNRs seen with $ROSAT$ as elongated plasma-filled superbubbles on 
$\sim$100-pc scales ({\em Wang and Helfand}, 1991). 

The new {\it Chandra} image shows a bright concentration of X-rays 
associated with the R136 star cluster, the bright SNR N157B to the 
southwest, a number of new widely-distributed compact X-ray sources, 
and diffuse structures which fill the superbubbles produced by the 
collective effects of massive stellar winds and their past supernova 
events ({\em Townsley et al.}, 2006a).  Some of these are 
center-filled while others are edge-brightened, indicating a 
complicated mix of viewing angles and perhaps filling factors. 
Comparison of the morphologies of the diffuse X-ray emission with the 
photodissociation region revealed by H$\alpha$ imaging and cool dust 
revealed by infrared imaging with the {\it Spitzer Space Telescope} 
shows a remarkable association: the hot plasma clearly fills the 
cavities outlined by ionized gas and warm dust. Spectral analysis of 
the superbubbles reveals a range of absorptions ($A_V = 1-3$ mag), 
plasma temperatures ($T = 3-9$ MK), and abundances.  About 100 X-ray 
sources are associated with the central massive cluster R136 ({\em 
Townsley et al.}, 2006b).  Some are bright, hard X-ray point sources 
in the field likely to be colliding-wind binaries, while others are 
probably from ordinary O and WR stellar winds.

\bigskip
{\textbf{ 5. X-RAY EFFECTS ON STAR AND PLANET FORMATION}}
\bigskip

\bigskip
\noindent \textbf{5.1 X-ray ionization of molecular cloud cores}
\bigskip

One of the mysteries of Galactic astrophysics is why most 
interstellar molecular material is not engaged in star formation.  
Large volumes of most molecular clouds are inactive, and some clouds 
appear to be completely quiescent.   A possible explanation is that 
star formation is suppressed by ionization: stellar ultraviolet will 
ionize the outer edges of clouds, and Galactic cosmic rays may 
penetrate into their cores ({\em Stahler and Palla}, 2005). Even very 
low levels of ionization will couple the mostly-neutral gas to 
magnetic fields, inhibiting gravitational collapse until sufficient 
ambipolar diffusion occurs. 

The X-ray observations of star forming regions demonstrate that a 
third source of ionization must be considered: X-rays from the winds 
and flares of deeply embedded X-ray sources. The X-ray ionization 
zones, sometimes called X-ray Dissociation Regions (XDRs) or 
R\"ontgen Spheres, do not have sharp edges like ultraviolet 
Str\"omgren Spheres, but rather extend to large distances with 
decreasing effect ({\em Hollenbach and Tielens}, 1997).  

The COUP observation provides a unique opportunity to calculate
realistic XDRs in two molecular cloud cores:  OMC-1 or the
Becklin-Neugebauer region, and OMC-1 South.  Several dozen embedded
X-ray stars are seen in these clouds (Plate 1), and each can be
characterized by X-ray luminosity, spectrum and line-of-sight
absorption ({\em Grosso et al.}, 2005).  Fig.\ \ref{protostar.fig}
illustrates the X-ray properties of deeply embedded objects.  COUP 554
is a young star with a strong infrared-excess in the OMC-1 South core.
The {\it Chandra} spectrum shows soft X-ray absorption of $\log N_H[cm^{-2}] =
22.7$, equivalent to $A_V \sim 30$ mag, and the lightcurve
exhibits many powerful flares at the top of the XLF with peak X-ray
luminosities reaching $\sim 10^{32}$ erg/s.  COUP 632 has no optical or
$K$-band counterpart and its X-ray spectrum shows the strongest
absorption of all COUP sources:  $\log N_H[cm^{-2}] \simeq 23.9$ or
$A_V \sim 500$ mag.

Using the COUP source positions and absorptions, we can roughly place
each star into a simplified geometrical model of the molecular cloud
gas, and calculate the region around each where the X-ray ionization
exceeds the expected uniform cosmic ray ionization.  Plate 2 shows the
resulting XDRs in OMC-1 from the embedded Becklin-Neugebauer cluster
({\em Lorenzani et al.}, in preparation).  Here about one-fourth of
the volume is dominated by X-ray ionization.  In general, the
ionization of cloud cores $\simeq 0.1$ pc in size will be
significantly altered if they contain clusters with more than $\sim
50$ members.

\begin{figure*}[ht]
 \epsscale{1.7} 
 \plotone{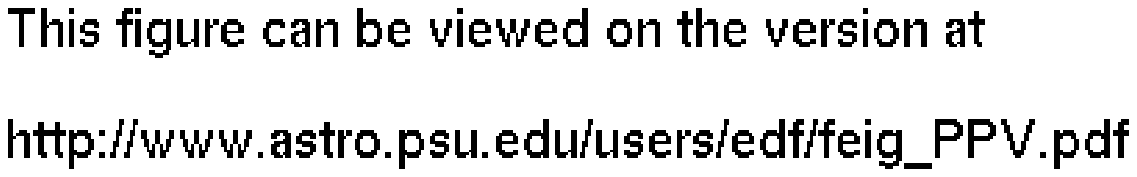} 
 \caption{\small  Cartoon illustrating sources of energetic irradiation 
(Galactic cosmic rays, flare X-rays, flare particles) and their 
possible effects on protoplanetary disks (ionization of gas and 
induction of MHD turbulence, layered accretion structure, spallation 
of solids).   ({\em Feigelson}, 2005) \label{disk_irr.fig}}
\end{figure*}

\bigskip
\noindent \textbf{5.2 X-ray irradiation of protoplanetary disks}
\bigskip

The circumstellar disks around PMS stars where planetary systems form
were generally considered to consist of cool, neutral molecular
material in thermodynamic equilibrium with $\sim 100-1000$ K
temperatures.  But there is a growing understanding that they are not
closed and isolated structures.  A few years ago, discussion
concentrated on ultraviolet radiation from O stars which can
photoevaporate nearby disks ({\em Hollenbach et al.}, 2000).  More
recently, considerable theoretical discussion has focused on X-ray
irradiation from the host star (Fig.\ \ref{disk_irr.fig}).  This is a
rapidly evolving field and only a fraction of the studies can be
mentioned here.  Readers are referred to reviews by {\em Feigelson}
(2005) and {\em Glassgold et al.}  (2005) for more detail.

X-ray studies provide two lines of empirical evidence that the X-rays 
seen with our telescopes actually do efficiently irradiate 
protoplanetary disks.  First, the 6.4 keV fluorescent line of neutral 
iron is seen in several embedded COUP stars with massive disks, as 
shown in Fig.\ \ref{YLW16a.fig} ({\em Imanishi et al.}, 2001).  This 
line is only produced when hard X-rays illuminate $>1$ g/cm$^2$ of 
material; this is too great to be intervening material and must be 
reflection off of a flattened structure surrounding the X-ray source 
{\em Tsujimoto et al.}, 2005; {\em Favata et al.}, 2005b).  Second, 
X-ray spectra of PMS stars with inclined disks show more absorption 
than spectra from stars with face-on disks.  This is most clearly 
seen in the COUP survey where column densities $\log N_H$[cm$^{-2}$]$ 
\sim 23$ are seen in edge-on proplyds imaged with the {\it Hubble 
Space Telescope}({\em Kastner et al.}, 2005).  This demonstrates the 
deposition of ionizing radiation in the disk and gives a rare 
measurement of the gas (rather than dust) content of protoplanetary 
disks.  

Having established that X-ray emission, particularly X-ray flaring, 
is ubiquitous in PMS stars, and that at least some disks are 
efficiently irradiated by these X-rays, one can now estimate the 
X-ray ionization rate throughout a disk. The result is that X-rays 
penetrate deeply towards the midplane in the Jovian planet region, 
but leave a neutral `dead zone' in the terrestrial planet region 
(e.g., {\em Igea and Glassgold}, 1999; {\em Fromang et al.}, 2002).  
The ionization effect of X-rays is many orders of magnitude more 
important than that of cosmic rays. However, differing treatments of 
metal ions and dust leads to considerable differences in the inferred 
steady-state ionization level of the disk {(\em Ilgner and Nelson}, 
2006). The theory of the X-ray ionization {\it rate} thus appears 
satisfactory, but calculations of the X-ray ionization {\it fraction} 
depend on poorly established recombination rates.

X-ray ionization effects become important contributors to the complex 
and nonlinear interplay between the thermodynamics, dynamics, 
gas-phase chemistry and gas-grain interactions in protoplanetary 
disks.  One important consequence may be the induction of the 
magnetorotational instability which quickly develops into a full 
spectrum of MHD turbulence including both vertical and radial mixing.  
The radial viscosity associated with the active turbulent zone may 
cause the flow of material from the outer disk into the inner disk, 
and thereby into the bipolar outflows and onto the protostar.  This 
may solve a long-standing problem in young stellar studies:  a 
completely neutral disk should have negligible viscosity and thus 
cannot efficiently be an accretion disk.  Ionization-induced 
turbulence should affect planet formation and early evolution in 
complex ways: suppressing gravitational instabilities, concentrating 
solids, producing density inhomogeneities which can inhibit Type I 
migration of protoplanets, diminishing disk gaps involved in Type II 
migration, and so forth.  It is thus possible that X-ray emission 
plays an important role in regulating the structure and dynamics of 
planetary systems, and the wide range in X-ray luminosities may be 
relevant to the diversity of extrasolar planetary systems.  

PMS X-rays are also a major source of ionization at the base of 
outflows from protostellar disks which produce the emission line 
Herbig-Haro objects and molecular bipolar outflows ({\em Shang et 
al.},  2002).  This is a profound result:  if low-mass PMS stars were 
not magnetically active and profusely emitting penetrating 
photoionizing radiation, then the coupling between the Keplerian 
orbits in the disk and the magnetocentrifugal orbits spiralling 
outward perpendicular to the disks might be much less efficient than 
we see.  

X-ray ionization of a molecular environment will induce a complex 
series of molecular-ion and radical chemical reactions (e.g., {\em 
Aikawa and Herbst}, 1999; {\em Semenov et al.}, 2004).  CN, HCO+ and 
C$_2$H abundances may be good tracers of photoionization effects, 
though it is often difficult to distinguish X-ray and ultraviolet 
irradiation from global disk observations.  X-ray heating may also 
lead to ice evaporation and enhanced gaseous abundances of molecules 
such as methanol. X-ray absorption also contributes to the warming of 
outer molecular layers of the disk.  In the outermost layer, the gas 
is heated to 5000 K, far above the equilibrium dust temperature ({\em 
Glassgold et al.}, 2004).  This may be responsible for the strong 
rovibrational CO and H$_2$ infrared bands seen from several young 
disks. 

Finally, PMS flaring may address several long-standing 
characteristics of ancient meteorites which are difficult to explain 
within the context of a quiescent solar nebula in thermodynamic 
equilibrium: \begin{enumerate}

\item Meteorites reveal an enormous quantity of flash-melted 
chondrules.  While many explanations have been proposed, many with 
little empirical support, it is possible that they were melted by the 
$>10^8$ X-ray flares experienced by a protoplanetary disk during the 
era of chondrule melting. Melting might either be produced directly 
by the absorption of X-rays by dustballs ({\em Shu et al.}, 2001), or 
by the passage of a shock along the outer disk ({\em Nakamoto et 
al.}, 2005).  

\item Certain meteoritic components, particularly the 
calcium-aluminum-rich inclusions (CAIs), exhibit high abundances of 
daughter nuclides of short-lived radioisotopic anomalies which must 
have been produced immediate before or during disk formation.  Some 
of these may arise from the injection of recently synthesized 
radionuclides from supernovae, but other may be produced by 
spallation from MeV particles associated with the X-ray flares ({\em 
Feigelson et al.}, 2002). Radio gyrosynchrotron studies already 
demonstrate that relativistic particles are frequently present in 
PMS systems.  

\item Some meteoritic grains that were free-floating in the solar
nebula show high abundances of spallogenic $^{21}$Ne excesses 
correlated with energetic particle track densities ({\em Woolum and 
Hohenberg}, 1993).  The only reasonable explanation is irradiation 
by high fluences of MeV particles from early solar flares. 
\end{enumerate}

We thus find that X-ray astronomical studies of PMS stars have a wide 
variety of potentially powerful effects on the physics, chemistry and 
mineralogy of protoplanetary disks and the environment of planet 
formation.  These investigations are still in early stages, and it is 
quite possible that some of these proposed effects may prove to be 
uninteresting while others prove to be important.

\bigskip
{\textbf{ 6.  SUMMARY}}
\bigskip

The fundamental result of X-ray studies of young stars and star 
formation regions is that material with characteristic energies of 
keV (or even MeV) per particle is present in environments where the 
equilibrium energies of the bulk material are meV.  Magnetic 
reconnection flares in lower mass PMS stars, and wind shocks on 
different scales in O stars, produce these hot gases.  Although the 
X-ray luminosities are relatively small, the radiation effectively 
penetrates and ionizes otherwise neutral molecular gases and may even 
melt solids. X-rays from PMS stars may thus have profound effects on 
the astrophysics of star and planet formation.

The recent investigations outlined here from the {\it Chandra} and 
{\it XMM-Newton} observatories paint a rich picture of X-ray emission 
in young stars. Both the ensemble statistics and the characteristics 
of individual X-ray flares strongly resemble the flaring seen in the 
Sun and other magnetically active stars. Astrophysical models of 
flare cooling developed for solar flares fit many PMS flares well. 
PMS spectra show the same abundance anomalies seen in older stars. 
Rotationally modulated X-ray variability of the non-flaring 
characteristic emission show that the X-ray emitting structures lie 
close to the stellar surface and are inhomogeneously distributed in 
longitude.  This is a solid indication that the X-ray emitting 
structures responsible for the observed modulation are in most cases 
multipolar magnetic fields as on the Sun.

At the same time, the analysis of the most powerful flares indicates 
that very long magnetic structures are likely present in some of the 
most active PMS stars, quite possibly connecting the star with its 
surrounding accretion disk.  The evidence suggests that both 
coronal-type and star-disk magnetic field lines are present in PMS 
systems, in agreement with current theoretical models of 
magnetically funnelled accretion.

There is a controversy over the X-ray spectra of a few of the 
brightest accreting PMS stars.  TW Hya shows low plasma temperatures 
and emission lines suggesting an origin in accretion shocks rather 
than coronal loops.  However, it is a challenge to explain the 
elemental abundances and to exclude the role of ultraviolet 
irradiation.  Simultaneous optical observations during the COUP X-ray 
observation clearly shows that the bulk of X-ray emission does not 
arise from accretion processes.  Perhaps counterintuitively, various 
studies clearly show that accreting PMS stars are statistically 
weaker X-ray emitters than non-accretors.  A fraction of the magnetic 
field lines in accreting PMS stars are likely to be mass loaded and 
can not reach X-ray temperatures.  

X-ray images of high-mass star forming regions are incredibly rich 
and complex.  Each image shows hundreds or thousands of magnetically 
active PMS stars with ages ranging from Class~1 (and controversially, 
Class 0) protostars to Zero-Age Main Sequence stars.  Hard X-rays are 
often emitted so that {\it Chandra} can penetrate up to $A_V \simeq 
500$ mag into molecular cloud material. {\it Chandra} images of MSFRs 
also clearly reveal for the first time the fate of O star winds: the 
interiors of some HII regions are suffused with a diffuse 10 MK 
plasma, restricting the $10^4$K gas to a thin shell.  The concept of 
a Str{\" o}mgren Sphere must be revised in these cases. Only a small 
portion of the wind energy and mass appears in the diffuse X-ray 
plasma; most likely flows unimpeded into the Galactic interstellar 
medium. The full population of stars down to $\sim 1$ M$_\odot$ is 
readily seen in X-ray images of MSFRs, with little contamination from 
extraneous populations.  This may lead, for example, to X-ray-based 
discrimination of close binaries with colliding winds and 
identification of intermediate-mass PMS stars that are not accreting.  
In the most active and long-lived MSFRs, cavity SNRs and superbubbles 
coexist with, and may dominate, the stellar and wind X-ray 
components.  X-ray studies thus chronicle the life cycle of massive 
stars from proto-O stars to colliding O winds, to supernova remnants 
and superbubbles. These star forming regions represent the building 
blocks of Galactic scale star formation and starburst galaxies.  

Some of the issues discussed here are now well-developed while others 
are still in early stages of investigation.  It is unlikely that 
foreseeable studies will give qualitatively new information on the 
X-ray properties of low-mass PMS stars than obtained from the many 
studies emerging from the COUP and XEST projects. In-depth analysis 
of individual objects, especially high-resolution spectroscopic 
study, represents an important area ripe for follow-up exploration.  
The many X-ray studies of MSFRs now emerging should give large new 
samples of intermediate-mass stars, and new insights into the complex 
physics of OB stellar winds on both small and large scales.  Although 
{\it Chandra} and {\it XMM-Newton} have relatively small fields, a 
commitment to wide-field mosaics of MSFR complex like W3-W4-W5 and 
Carina could give unique views into the interactions of high-mass 
stars and the Galactic interstellar medium.  Deep X-ray exposures are 
needed to penetrate deeply to study the youngest embedded systems.  
Finally, the next generation of high-throughput X-ray telescopes 
should bring new capabilities to perform high-resolution spectroscopy 
of the X-ray emitting plasmas. Today, theoretical work is urgently 
needed on a host of issues raised by the X-ray findings: magnetic 
dynamos in convective stars, accretion and reconnection in disk-star 
magnetic fields, flare physics at levels far above those seen in the 
Sun, and possible effects of X-ray ionization of protoplanetary 
disks.

\textbf{~ Acknowledgments.} EDF recognizes the excellent work by 
Konstantin Getman and the other 36 scientists in the COUP 
collaboration.  EDF and LKT benefit from discussions with their Penn 
State colleagues Patrick Broos, Gordon Garmire, Konstantin Getman, 
Masahiro Tsujimoto, and Junfeng Wang.  Penn State work is supported 
by the National Aeronautics and Space Administration (NASA) through 
contract NAS8-38252 and {\em Chandra} Awards G04-5006X, G05-6143X, 
and SV4-74018 issued by the {\em Chandra} X-ray Observatory Center, 
operated by the Smithsonian Astrophysical Observatory for and on 
behalf of NASA under contract NAS8-03060. MG warmly acknowledges the 
extensive work performed by XEST team members.  The XEST team has 
been financially supported by the Space Science Institute (ISSI) in 
Bern, Switzerland. {\it XMM-Newton} is an ESA science mission with 
instruments and contributions directly funded by ESA Member States 
and the USA (NASA).  KGS is grateful for funding support from NSF 
CAREER grant AST-0349075.

\bigskip

\centerline\textbf{ REFERENCES}
\bigskip
\parskip=0pt
{\small \baselineskip=11pt 

\refs Aikawa Y. and Herbst E.\ (1999) {\it Astron. Astrophys., 351}, 
233-246.

\refs Albacete Colombo J.~F., M\'endez M.\ and Morrell N.~I.\ (2003) 
{\it Mon. Not. Roy. Astr. Soc. 346}, 704-718.

\refs Arzner K. and G{\"u}del M.\ (2004) {\em Astrophys. J., 602}, 
363-376.
 
\refs Audard M., et al.\ (2005) {\it Astrophys. J., 635}, L81-L84.

\refs Barnes S.~A.\ (2003a) {\it Astrophys. J., 586}, 464-479. 

\refs Barnes S.~A.\ (2003b) {\it Astrophys. J., 586}, L145-L147.  

\refs Brinkman A. C., Behar E., G{\"u}del M., Audard M., den Boggende 
A. J. F., et al. (2001) {\it Astron. Astrophys., 365}, L324-L328.

\refs Calvet N., et al.\ (2002) {\it Astrophys. J., 568}, 1008-1016.  

\refs Cant\'{o} J., Raga A.~C., and Rodr\'{\i}guez L.~F.\ (2000) {\it 
Astrophys. J., 536}, 896-901. 

\refs Capriotti E.~R. and Kozminski J.~F.\ (2001) {\it Publ. Astron. 
Soc. Pac., 113}, 677-691.

\refs Chu Y.-H., Mac Low M.-M., Garcia-Segura G., Wakker B., and 
Kennicutt, R. C.\ (1993) {\it Astrophys. J., 414}, 213-218. 

\refs Damiani F., Flaccomio E., Micela G., Sciortino S., Harnden 
F.~R.~Jr.\ and Murray S.~S.\ (2004) {\it Astrophys. J. 608}, 781-796.

\refs Drake J.~J., Testa P., and Hartmann L.\ (2005) {\it Astrophys. 
J., 627}, L149-L152. 

\refs Dunne B. C., et al.\ (2003) {\it Astrophys. J., 590}, 306-313.

\refs Evans N.~R., et al.\ (2003) {\it Astrophys. J. 589}, 509-525. 

\refs Evans N.~R., et al.\ (2004) {\it Astrophys. J. 612}, 1065-1080. 

\refs Ezoe Y., Kokubun M., Makishima K., Sekimoto Y.\ and Matsuzaki 
K.\ (2006) {\it Astrophys. J.}, in press.

\refs Favata F.\ and Micela, G.\ (2003) {\it Space Sci. Rev., 108}, 
577-708.

\refs Favata F., Giardino G., Micela G., Sciortino S., and Damiani 
F.\ (2003) {\it Astron. Astrophys., 403}, 187-203. 

\refs Favata F., et al.\  (2005a) {\it  Astrophys. J. Suppl., 160}, 
469-502. 

\refs Favata F., Micela G., Silva B., Sciortino S.\ and Tsujimoto M.\ 
(2005b) {\it Astron. Astrophys. 433}, 1047-1054.

\refs Feigelson, E.~D.\ (2005) In {\it Proc.\ 13th Cool Stars 
Workshop}, eds.\ F.\ Favata et al. (ESA, SP-560), pp.\ 175-183.

\refs Feigelson E.~D. and Montmerle T.\ (1999) {\it Ann. Rev.  
Astron. Astrophys., 37}, 363-408. 

\refs Feigelson E.~D., Garmire G.~P., and Pravdo S.~H.\ (2002) {\it 
Astrophys. J., 572}, 335-349.

\refs Feigelson E.~D., Gaffney J.~A., Garmire G., Hillenbrand L.~A., 
and Townsley L.\ (2003) {\it Astrophys. J., 584}, 911-930. 

\refs Feigelson E. D., et al.\ (2004) {\it Astrophys. J., 611}, 
1107-1120. 

\refs Feigelson E. D., et al.\ (2005) {\it Astrophys. J. Suppl., 160}, 
379-389. 

\refs Flaccomio E., Micela G., and Sciortino S.\ (2003) {\it Astron. 
Astrophys., 402}, 277-292. 

\refs Flaccomio E., et al.\ (2005) {\it Astrophys. J. Suppl., 160}, 
450-468.  

\refs Fromang S., Terquem C.\ and Balbus, S.~A.\ (2002) {\it Mon. 
Not. Roy. Astr. Soc. 329}, 18-28. 

\refs Gagn\'e M., Skinner S.~L., and Daniel K.~J.\ (2004) {\it 
Astrophys. J., 613}, 393-415.

\refs Getman K.~V., Feigelson E.~D., Townsley L., Bally J., Lada 
C.~J., and Reipurth B.\ (2002) {\it Astrophys. J., 575}, 354-377.  

\refs Getman K. V., et al.\ (2005a) {\it Astrophys. J. Suppl., 160}, 
319-352.

\refs Getman K.~V., et al.\ (2005b) {\it Astrophys. J. Suppl., 160}, 
353-378.

\refs Getman K. V., et al.\ (2006) {\it Astrophys. J. Suppl.}, in 
press.

\refs Glassgold A.~E., Feigelson E.~D., and Montmerle T.\ (2000)  In 
{\it Protostars and Planets IV}, eds.\ V. Mannings et al.\ (Tucson: Univ.\
Arizona Press), pp.\ 429-456.

\refs Glassgold A.~E., Najita J., and Igea J.\ (2004) {\it Astrophys. 
J. 615}, 972-990.

\refs Glassgold A.~E., Feigelson E.~D., Montmerle T.\ and Wolk S.\ 
(2005) In {\it Chondrites and the Protoplanetary Disk}, eds. 
A.~N. Krot et al., (San Francisco: ASP Conf.\ Ser.\ 341), pp. 161-180.

\refs Grosso N., Montmerle T., Feigelson E.~D., and Forbes T.~G.\ 
(2004) {\it Astron. Astrophys., 419}, 653-665. 

\refs Grosso N., et al.\ (2005) {\it Astrophys. J. Suppl. 160}, 
530-556.

\refs G{\"u}del M.\ (2002) {\it Ann. Rev. Astron. Astrophys., 40}, 
217-261. 

\refs G{\"u}del M.\ (2004) {\it Astron. Astrophys. Rev., 12}, 71-237.  

\refs G{\"u}del M., Guinan E.~F., and Skinner S.~L.\ (1997) {\it 
Astrophys. J., 483}, 947-960.

\refs G\"udel M., Audard M., Kashyap V.~L., Drake J.~J., and Guinan 
E.~F. (2003) {\em Astrophys. J., 582}, 423-442.

\refs Hamaguchi K., et al.\ (2005) {\it Astrophys. J., 623}, 291-301.

\refs Hartmann L.\ (1998) {\it Accretion processes in star 
formation}, Cambridge Univ.\ Press, New York.

\refs Herbst W., Bailer-Jones C.~A.~L., Mundt R., Meisenheimer K., 
and Wackermann R.\ (2002) {\it Astron. Astrophys., 396}, 513-532. 

\refs Hollenbach, D. J., Yorke, H. W., Johnstone, D.\ (2000) In {\it 
Protostars and Planets IV} , eds. V.\ Mannings et al.\ (Tucson: Univ.\
Arizona Press), pp. 401-416.

\refs Hollenbach, D.~J.\ and Tielens, A.~G.~G.~M.\ (1997) {\it Ann. 
Rev. Astron. Astrophys., 35}, 179-216.

\refs Igea J., and Glassgold A.~E.\ (1999) {\it Astrophys. J., 518}, 
848-858. 

\refs Imanishi K., Koyama K., and Tsuboi Y.\ (2001) {\it Astrophys. 
J., 557}, 747-760.  

\refs Ilgner M., and Nelson R.~P.\ (2006) {\it Astron. Astrophys., 
445}, 223-232.

\refs Jardine M. and Unruh Y.~C.\ (1999) {\it Astron. Astrophys., 
346}, 883-891. 

\refs Jardine M, Collier Cameron A., Donati J.-F., Gregory S.~G. and 
Wood K.\ (2006) {\it Mon. Not. Roy. Astr. Soc.}, in press.

\refs Johns-Krull C.~M., Valenti J.~A., and Saar S.~H.\ (2004) {\it 
Astrophys. J., 617}, 1204-1215.

\refs Kahn F.~D.~and Breitschwerdt D.\ (1990) {\it Mon. Not. R. 
Astron. Soc., 242}, 209-214.

\refs Kashyap V.~L., Drake J.~J., G{\"u}del M., and Audard M.\ (2002) 
{\it Astrophys. J., 580}, 1118-1132. 

\refs Kastner J.~H., et al.\ (2004), {\it Nature, 430}, 429-431. 

\refs Kastner J.~H., Huenemoerder D.~P., Schulz N.~S., Canizares 
C.~R., and Weintraub D.~A.\ (2002) {\it Astrophys. J., 567}, 434-440. 

\refs Kastner J.~H., et al.\ (2005) {\it  Astrophys. J. Suppl., 160}, 
511-529.

\refs Lamzin S.~A.\ (1999) {\it Astron. Letters, 25}, 430-436.

\refs Law, C.\ and Yusef-Zadeh F.\ (2004) {\it Astrophys. J. 611}, 
858-870.

\refs Loinard L., et al.\ (2005) {\it Astrophys. J., 619}, L179-L182. 

\refs Montmerle T., Grosso N., Tsuboi Y., and Koyama K.\ (2000) {\it 
Astrophys. J., 532}, 1097-1110.  

\refs Mullan D.~J. and MacDonald J.\ (2001) {\it Astrophys. J., 559}, 
353-371.

\refs Muno M.~P.\ et al.\ (2006) {\it Astrophys. J. 636}, L41-L44.

\refs Nakamoto T., and Miura H. In {\it PPV Poster Proceedings} \\ 
http://www.lpi.usra.edu/meetings/ppv2005/pdf/8530.pdf

\refs Nelan E. P., et al.\ (2004) {\it Astron. J., 128}, 323-329.

\refs Ness J.-U. and Schmitt J.~H.~M.~M.\ (2006) {\it Astron. 
Astrophys.}, in press.

\refs Nielbock M., Chini R., and M{\" u}ller S.~A.~H.\ (2003) {\it 
Astron. Astrophys., 408}, 245-256.

\refs Pace G. and Pasquini L.\ (2004) {\it Astron. Astrophys., 426}, 
1021-1034.

\refs Paerels F.~B.~S.\ and Kahn S.~M.\ (2003) {\it Ann. Rev. Astron. 
Astrophys., 41}, 291-342.

\refs Parker, E.~N.\ (1998) {\it Astrophys. J. 330}, 474-479.

\refs Pittard J.~M., Hartquist T.~W., and Dyson J.~E.\ (2001) {\it 
Astron. Astrophys., 373}, 1043-1055.

\refs Preibisch, T., Neuh\"auser, R., and Alcal\'a, J.~M.\ (1995)
{\it Astron. Astrophys., 304}, L13-L16.

\refs Preibisch T.\ (2004) {\it Astron. Astrophys., 428}, 569-577. 

\refs Preibisch T. and Feigelson E.~D.\ (2005) {\it Astrophys. J. 
Suppl., 160}, 390-400. 

\refs Preibisch T., et al.\ (2005) {\it Astrophys. J. Suppl., 160}, 
401-422. 

\refs Priest E.~R.\ and Forbes T.~G.\ (2002) {\it Astron. Astrophys. 
Rev, 10}, 313-377.

\refs Rockefeller G., Fryer C.~L., Melia F.\ and Wang, Q.~D.\ (2005) 
{\it Astrophys. J. 623}, 171-180.

\refs Scelsi L.,  Maggio A., Peres G., and Pallavicini R. (2005) {\it 
Astron. Astrophys., 432}, 671-685.

\refs Schmitt J.~H.~M.~M., Robrade J., Ness J.-U., Favata F., and 
Stelzer B.\ (2005) {\it Astron. Astrophys., 432}, L35-L38. 

\refs Schrijver C.~J.\ and Zwaan C.\ (2000) {\it Solar and Stellar 
Magnetic Activity}, Cambridge Univ.\ Press, New York.

\refs Semenov D., Weibe D.\ and Henning, Th.\ (2004) {\it Astron. 
Astrophys. 417}, 93-106.

\refs Seward F.~D. and Chlebowski T.\ (1982) {\it Astrophys. J., 
256}, 530-542.

\refs Shang H., Glassgold A.~E., Shu F.~H., Lizano S. (2002) {\it 
Astrophys. J., 564}, 853-876. 

\refs Shu F.~H., Najita J.~R., Shang H., and Li Z.-Y.\ (2000) In {\it 
Protostars and Planets IV}, eds.\ V.\ Mannings et al., Tucson: Univ.\
Arizona Press), pp.\ 789-814,.

\refs Shu F.~H., Shang H., Gounelle M., Glassgold A.~E., and Lee T.\ 
(2001) {\it Astrophys. J., 548}, 1029-1050.  
 
\refs Siess L., Forestini M., and Bertout C.\ (1999) {\it Astron. 
Astrophys., 342}, 480-491. 

\refs Skinner S., Gagn\'e M.\ and Belzer E.\ (2003) {\it Astrophys. 
J. 598}, 375-391.

\refs Skinner S.~L., Zhekov S.~A., Palla F.\ and Barbosa C.~L.~D.~R. 
(2005) {\it Mon. Not. Roy. Astr. Soc. 361}, 191-205.

\refs Skinner S.~L., et al.\ (2006) {\it Astrophys. J. Lett.}, in 
press 

\refs Skumanich A.\ (1972) {\it Astrophys. J., 171}, 565-567. 

\refs Stahler, S.~W.\ and Palla, F.\ (2005) {\it The Formation of 
Stars}, Wiley-VCH, New York.

\refs Stassun K.~G., Ardila D.~R., Barsony M., Basri G., and Mathieu 
R.~D.\ (2004) {\it Astron. J., 127}, 3537-3552.  

\refs Stelzer B. and Schmitt J.~H.~M.~M.\ (2004) {\it Astron. 
Astrophys., 418}, 687-697. 

\refs Swartz, D.~A., et al.\ (2005) {\it Astrophys. J., 628}, 811-816.

\refs Townsley L.~K., et al.\ (2003) {\it Astrophys. J., 593}, 
874-905.

\refs Townsley L.~K., et al.\ (2006a) {\it Astron. J.}, in press.

\refs Townsley L.~K., Broos P.~S., Feigelson E.~D., Garmire G.~P., 
and Getman K.~V.\ (2006b) {\it Astron. J.}, in press.

\refs Tsuboi Y., Imanishi K., Koyama K., Grosso N., and Montmerle T.\ 
(2000) {\it Astrophys. J., 532}, 1089-1096.  

\refs Tsuboi Y., et al.\ (2001) {\it Astrophys. J., 554}, 734-741. 

\refs Tsujimoto, M., et al.\ (2005) {\it Astrophys. J. Suppl. 160}, 
503-510.

\refs Wang Q. and Helfand D.~J.\ (1991) {\it Astrophys. J., 373}, 
497-508.  

\refs Weaver R., McCray R., Castor J., Shapiro P., and Moore R.\ 
(1977) {\it Astrophys. J., 218}, 377-395.

\refs Wolk S.~J., et al.\ (2005) {\it Astrophys. J. Suppl., 160}, 
423-449. 

\refs Woolum D.~S. and Hohenberg C.\ (1993)  In {\it Protostars and 
Planets III}, eds.\ E.~H.\ Levy and J.~I.\ Lunine (Tuscon: Univ.\
Arizona Press), pp.\ 903-919.  

\newpage

\begin{figure*}
 \epsscale{1.4}
 \plotone{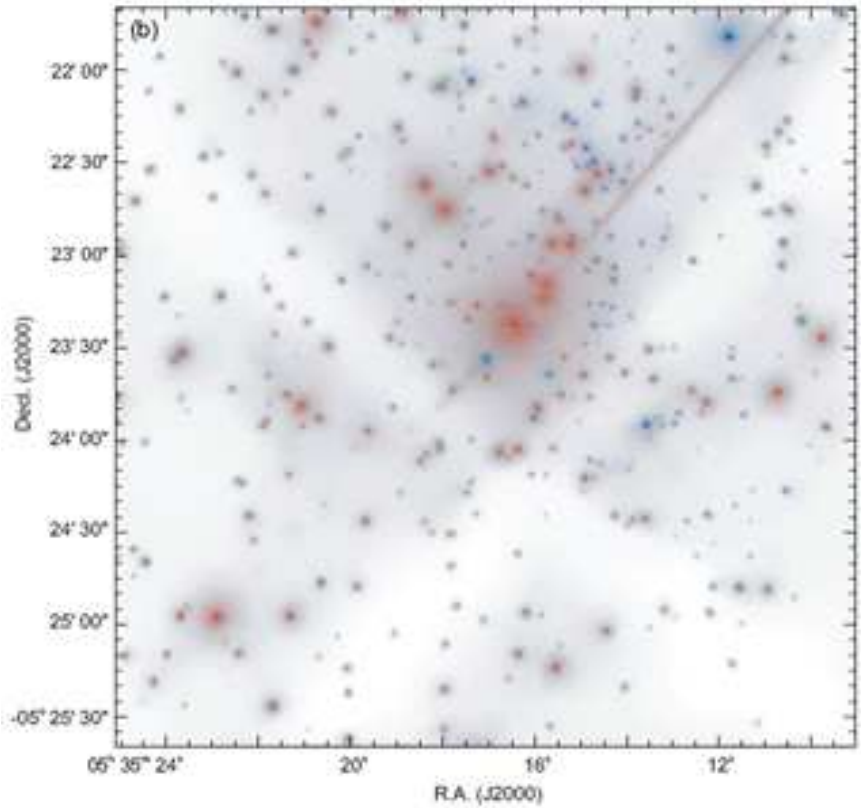}
 \caption{\small Plate 1.  Inner region of the Orion Nebula viewed by the {\it 
Chandra} Orion Ultradeep Project (COUP).  The image is smoothed and 
colors represent soft (red, $0.5-2$ keV) and hard (blue, $2-8$ keV) 
X-rays.  The brightest source is $\theta^1$C Orion (O7), and a group 
of embedded sources in the Becklin-Neugebauer region can be seen 1' 
to the northwest. ({\it Getman et al.}, 2005a) }
 \end{figure*}

\begin{figure*}
 \epsscale{1.0}
 \plotone{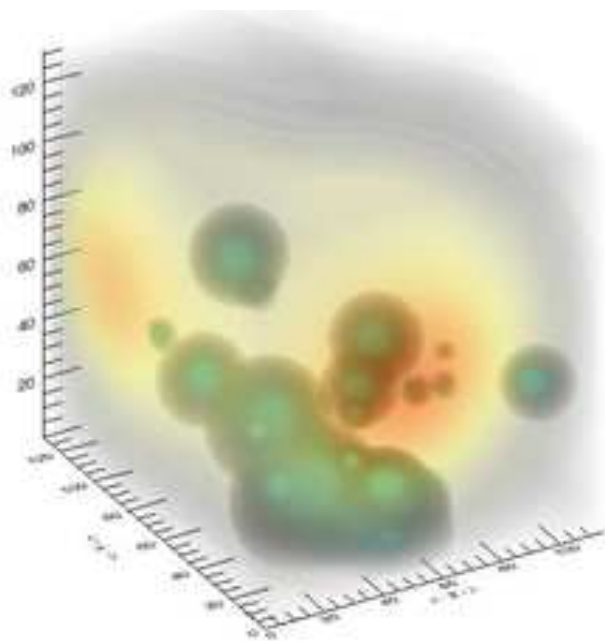}
\caption{\small Plate 2. X-ray dissociation regions from X-ray stars 
embedded in the Orion Becklin-Neugebauer cloud core. Yellow-to-red 
shows assumed three-dimensional structure of the molecular gas.  
Blue-to-green shows the inferred XDR structures.  ({\it Lorenzani et 
al.}, in preparation)  }
\end{figure*}

\newpage

\begin{figure*}
 \epsscale{2.2}
 \plottwo{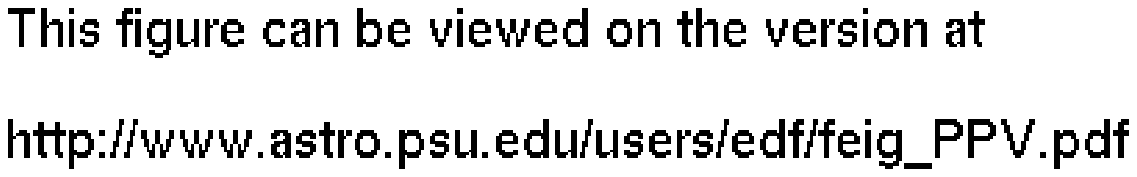}{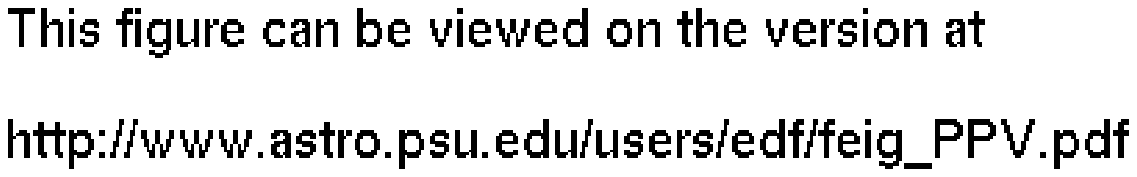}
 \caption{\small Plate 3. (a)  Smoothed {\it ROSAT} PSPC (soft X-ray) image of M17, $\sim 
39^{\prime} \times 42^{\prime}$, with the outline of the {\it 
Chandra} pointing overlaid in blue. (b)  The {\it Chandra} 
observation of M 17 showing hundreds of PMS stars and the soft X-ray 
flow from shocked O star winds.  Red intensity is scaled to the soft 
(0.5--2~keV) emission and blue intensity is scaled to the hard 
(2--8~keV) emission. ({\it Townsley et al.}, 2003) } \label{fig:m17}
\end{figure*}

\begin{figure*}
 \epsscale{2.2}
 \plottwo{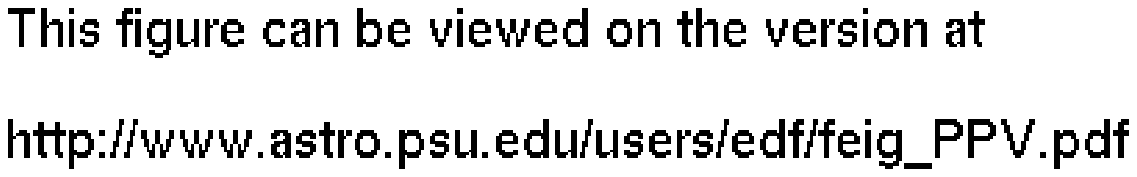}{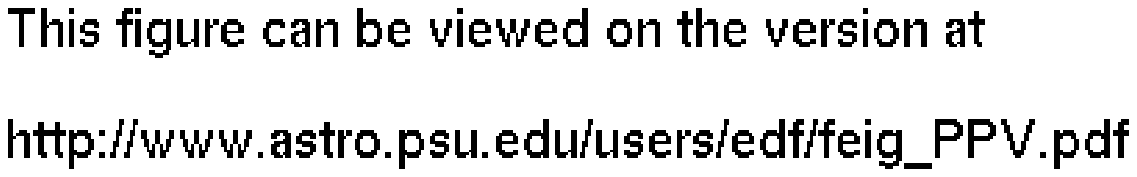}
\caption{Plate 4. (a)  A smoothed {\it Chandra} image (red = 
0.5--2~keV, blue = 2--8~keV) of Trumpler~14 in the Carina Nebula.  
The field is $17^\prime \times 17^\prime$, or $\sim 14 \times 14$~pc. 
About 1600 point sources plus extensive diffuse emission is seen.
({\it Townsley et al.}, in preparation) (b) Smoothed soft-band image 
(red = 0.5--0.7~keV, green = 0.7--1.1~keV, blue = 1.1--2.3~keV) of 
the 30~Doradus starburst in the Large Magellanic Cloud. The image 
covers $\sim 250$~pc on a side. R136a lies at the center; the 
plerionic SNR N157B lies to the southwest; the superbubble 30~Dor~C 
and the Honeycomb and SN1987A SNRs are seen in the two off-axis CCDs.  
The colorful inhomogeneous diffuse structures are the superbubbles 
produced by past generations of OB star winds and supernovae. ({\it 
Townsley et al.}, 2005a) } \label{fig:cavity}
\end{figure*}

\end{document}